%% file: main_hexValidity.tex
\documentclass[3p,times,procedia,number]{elsarticle}
\usepackage{ecrc}
\usepackage{amsmath}

\volume{00}

\firstpage{1}

\journalname{Procedia Engineering}

\runauth{A. Johnen et al.}

\jid{proeng}

\usepackage{amssymb}
\usepackage{amsthm}
\usepackage{bbold} 
\usepackage{arydshln} 
\usepackage{multirow}

\makeatletter
\DeclareRobustCommand{\sfrac}[1]{\@ifnextchar/{\@sfrac{#1}}%
                                            {\@sfrac{#1}/}}
\def\@sfrac#1/#2{\leavevmode\kern.1em\raise.5ex
         \hbox{$\m@th\mbox{\fontsize\sf@size\z@
                           \selectfont#1}$}\kern-.1em
         /\kern-.15em\lower.25ex
          \hbox{$\m@th\mbox{\fontsize\sf@size\z@
                            \selectfont#2}$}}
\makeatother

 \biboptions{sort&compress}

\usepackage[figuresright]{rotating}
\usepackage{bm}
\usepackage{tikz,pgfplots}
\usepackage{standalone}
\usetikzlibrary{calc}
\newlength\figwidth
\newlength\figheight
\graphicspath{{figures/hexes/}}
\usepackage{gensymb} 

\DeclareMathOperator\vol{vol}

\usepackage{xspace} 
\usepackage[
hyperfootnotes=true]{hyperref}
\usepackage{booktabs}
\newcommand{\ie}{\emph{i.e.}\xspace}
\newcommand{\eg}{\emph{e.g.}\xspace}

\newcommand{\beqn}[1]{\begin{equation}\label{#1}}
\newcommand{\eeqn}{\end{equation}}
\newcommand{\Par}[1]{\left(#1\right)}
\newcommand{\Cro}[1]{\left[#1\right]}
\renewcommand{\vec}[1]{\bm{#1}}
\newcommand\R{\mathbb R}

\newcommand\nchoosek[2]{{{#2}\choose {#1}}} 

\renewcommand{\o}{n} 
\newcommand\n[1]{\vec n_{#1}} 
\newcommand{\Np}{N} 
\newcommand\dom{\Omega} 
\newcommand\domref{{\dom_{\text{ref}}}} 

\newcommand\xra{\xi} 
\newcommand\xrb{\eta} 
\newcommand\xrc{\zeta} 
\newcommand\sx{x} 
\newcommand\x{\vec\sx} 
\newcommand\sxr{\xi} 
\newcommand\xr{\vec\sxr} 

\newcommand\zeros[2]{\mathbb0_{#1\times#2}} 
\newcommand\ones[2]{\mathbb1_{#1\times#2}} 
\newcommand\identity[2]{\mathbb I_{#1\times#2}} 
\newcommand\mf[2]{\sfrac{#1}{#2}} 

\newcommand\Tb{\vec T} 
\newcommand\Tsub{\vec T_{20\times20}} 
\newcommand\Tsmall{{\vec Q}} 

\newcommand\xx[2]{x_{#1#2}} 
\newcommand\yy[2]{y_{#1#2}} 
\renewcommand\v[2]{\vec v_{#1#2}} 

\renewcommand\d[2]{{#1}_{,#2}} 
\newcommand\da[1]{\d{#1}\xra} 
\newcommand\db[1]{\d{#1}\xrb} 
\newcommand\dc[1]{\d{#1}\xrc} 
\newcommand\Da[1]{\d{\Par{#1}}\xra} 
\newcommand\Db[1]{\d{\Par{#1}}\xrb} 
\newcommand\Dc[1]{\d{\Par{#1}}\xrc} 

\newcommand\jac{\vec J} 

\let\olddet\det
\renewcommand\det[1]{\olddet\Par{#1}} 
\newcommand\detJ{J} 

\newcommand\triple[3]{\Par{#1\times#2}\cdot#3}

\renewcommand{\L}[2]{L^{#1}_{#2}} 
\newcommand{\B}[2]{B^{#1}_{#2}} 
\renewcommand\i{{\alpha}} 
\renewcommand\j{{\beta}} 

\newtheorem{lemma}{Lemma}

\newtheorem{corollary}[lemma]{Corollary}

\newtheorem{observation}[lemma]{Observation}

\let\oldcite\cite
\def\cite{\,\oldcite}

\begin{document}

\begin{frontmatter}



\dochead{26th International Meshing Roundtable, IMR26, 18-21 September 2017, Barcelona, Spain}

\title{Robust and efficient validation of the linear hexahedral element}


\author[a]{A. Johnen}
\author[b]{J.-C. Weill}
\author[a]{J.-F. Remacle}

\address[a]{Universit\'e catholique de Louvain, Institute of Mechanics, 
Materials and Civil Engineering (iMMC), 
Avenue Georges Lemaitre 4, 1348 Louvain-la-Neuve, Belgium}
\address[b]{CEA, DAM, DIF, F-91297 Arpajon, France}

\begin{abstract}
Checking mesh validity is a mandatory step before doing 
any finite element analysis. If checking the validity of tetrahedra is
trivial, checking the validity of hexahedral elements is far from being obvious.
In this paper, a method that robustly and efficiently compute
the validity of standard linear hexahedral elements is presented. 
This method is a significant improvement of a previous work on the validity of curvilinear 
elements\cite{johnen2013geometrical}. The new implementation is
simple and computationally efficient. 
The key of the algorithm is still to compute B\'ezier coefficients of the 
Jacobian determinant. We show that only $20$ Jacobian determinants are necessary to 
compute the $27$ B\'ezier coefficients. Those $20$ Jacobians can be
efficiently computed by calculating the volume of $20$ tetrahedra. 
The new implementation is able to check the validity of about $6$ million
hexahedra per second on one core of a personal computer.
Through the paper, all the necessary information is provided that 
allow to easily reproduce the results, \ie write a simple code that
takes the coordinates of $8$ points as input and outputs the validity
of the hexahedron.
\end{abstract}


\end{frontmatter}

\email{amaury.johnen@uclouvain.be}



\section{Introduction}\label{s:intro}

Hexahedral meshes are often preferred to tetrahedral meshes as they offer 
excellent numerical properties: faster assembly\cite{remacle2016gpu}, orthogonal 
grids in the wall-normal direction for wall-bounded
flows, high accuracy in solid mechanics,
both for statics\cite{wang2004back} and dynamics,
or for quasi-incompressible materials\cite{benzley1995comparison}
\footnote{In many references, the
  accuracy of linear hexahedra is shown to be equivalent to the
  accuracy of quadratic tetrahedra with the same mesh density. Note
  that quadratic tetrahedra have one extra node per edge of the mesh,
  which multiplies the number of degrees of freedom by $7$.}.
Generating hex-meshes is however still an open problem for general 3D domains. 
Finite element meshes should ideally fill the 3D domain in a conformal
fashion but should also respect some size and quality constraints in
order to be suitable for finite element formulations. 
The validity of elements is usually the most important constraint and can be checked 
by verifying the local injectivity of their mapping; in the usual finite 
element language, one should check the positivity of the Jacobian determinant. 
While checking the validity of a linear tetrahedron just consists in ensuring 
its volume positivity, checking the validity of a linear hexahedron is not trivial.

Testing hexahedron validity is of particular interest when generating hex meshes
with an indirect method\cite{baudouin2014frontal,botella2016indirect,%
sokolov2016hexahedral}. In these methods, a huge set of hexahedral elements whose
cardinality can be as high as 40 times the number of vertices of the mesh is 
computed\cite{pellerin2017}. Computing the validity robustly and 
rapidly is then essential for the efficiency of these methods. 
Many algorithms have been proposed in the literature for checking the validity of 
hexahedra, however they do not provide 
any strong guarantees except method of\cite{johnen2013geometrical}. 
In this paper, we particularize this method for the linear hexahedron and propose
an efficient and simple implementation.

\paragraph{Previous works} 
\citet{knupp1990invertibility} has shown that the positivity of the 
Jacobian determinant at the $8$ corners of a linear hexahedron, 
as well as on its edges, is not sufficient to ensure its validity. 
He conjectured that any hexahedra having a positive Jacobian determinant on its 
boundary is valid. 
However, the Jacobian determinant on the faces are biquadratic functions; 
verifying their positivity is complex and, to our knowledge, no practical 
algorithm has been presented.


Some authors have proposed to check the validity by ensuring the
positivity of sets of tetrahedra constructed from the $8$ nodes of the
hexahedron\cite{ivanenko1999harmonic,grandy1999conservative,ushakova2001conditions,vavasis2003bernstein,shangyou2005subtetrabedral}. 
The number of tetrahedra ranges from $8$ to $64$.
\citet{ushakova2011nondegeneracy} compiled and empirically studied these tests. 
It is known that the positivity of the $8$ corner tetrahedra is a necessary
condition\cite{knupp1990invertibility,ivanenko1999harmonic}.
\citet{ushakova2011nondegeneracy} showed that none of the tests that consider less 
than $58$ tetrahedral volumes constitute a sufficient condition. 
The volume of the hexahedron is sometimes used in commercial
packages\cite{ushakova2001conditions}. It can be expressed from the volume of $10$ 
tetrahedra. It is a poor test alone but gives a sharper necessary condition when 
combined with the $8$ corner tetrahedra.

Another original method for checking the validity of linear hexahedra has been 
proposed by \citet{knabner2003conditions}. 
The Jacobian determinant of the hexahedron is expanded into the monomial basis.
Positivity conditions are derived from the monomial coefficients of respectively
a quadratic one-dimensional polynomial, a biquadratic polynomial and a triquadratic 
polynomial. The latter enables to check the positivity of the Jacobian determinant 
of the hexahedron. 
However, it is needed to linearize inequalities containing a square root which 
implies this approach to be only a sufficient condition.
A parameter provided by the user allows to determine the precision of this linearization. 

A method for checking the validity of curved finite element of any type has 
been proposed by \citet{johnen2013geometrical}. 
This method consists in expanding the Jacobian determinant into the B\'ezier basis 
of order 2. Thanks to the convex hull property of B\'ezier expansion, the minimum of 
these coefficients gives a lower bound of the Jacobian determinant. Moreover, the 
minimum of specific coefficients gives an upper bound of the minimum of the 
Jacobian determinant. These bounds are subsequently sharpened by ``subdividing'' 
in a recursive and adaptive manner which allows to compute the minimum of the 
Jacobian determinant with any prescribed tolerance. 
This method can be employed for the validity of the linear hexahedron since 
it is a particular case of the curved hexahedron. 

\paragraph{Contribution} 
To the best of our knowledge, the method\cite{johnen2013geometrical} is the 
only method to robustly check the validity of linear hexahedra. However, 
the general framework used for curved elements is not well-adapted for an efficient 
computation of the validity of one specific type of element. In this work, this 
method is optimized for to the specific case of the linear hexahedron. 
We start by introducing the validity of the linear quadrangle and hexahedron 
(\S\ref{s:validity}), and the B\'ezier expansion of the Jacobian determinant
(\S\ref{s:bezier}). Then, two substantial improvements are presented: 
we show that only $20$ quantities have to be computed 
instead of $27$ (\S\ref{s:only20}) and that those quantities can be computed as the 
volume of tetrahedra (\S\ref{s:tet2lag}). 
Finally, we present the complete algorithm (\S\ref{s:algo}) and demonstrate that 
this new algorithm is robust and efficient (\S\ref{s:results}). 
The C++ code implementing the algorithm will be available in 
Gmsh\cite{geuzaine2009gmsh} (\url{www.gmsh.info}).

\section{Validity of finite elements}\label{s:validity}

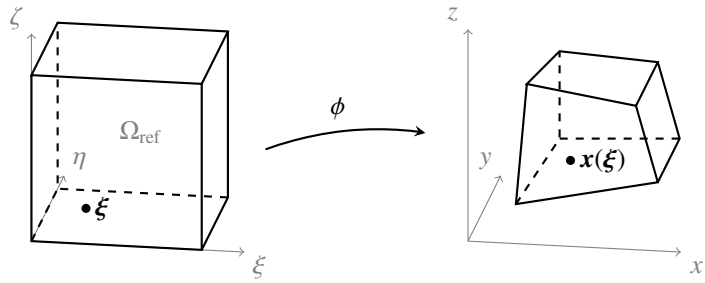
\begin{figure}[ht!]
\begin{center}
\setlength\figwidth{.5\textwidth}
\setlength\figheight{.3\textwidth}
\input{figures/tikz/mapping.tex}
\caption{Mapping between the reference and the physical hexahedron.}
\label{f:mapping}
\end{center}
\end{figure}

Let us consider a $d$-dimensional physical linear finite element which is 
geometrically defined by a set of
$N$ points $\n k\in\R^d,\ k = 1, \dots , \Np$, called nodes, and a set of Lagrange shape
functions $\L{}k(\xr): \domref\subset\R^d\to\R,\ k = 1, \dots , \Np$. These polynomial 
functions allow to map a reference unit element, represented by the domain of
definition $\domref$, to the physical element (see Figure~\ref{f:mapping}):
\beqn{e:mapping}
  \x(\xr) = \sum_{k=1}^{\Np}\L{}{k}(\xr)\,\n k. 
\eeqn

The Jacobian matrix of this mapping, denoted 
$\jac: \domref \to \R^{d\times d} : \xr \mapsto \jac(\xr)$, 
is by definition the matrix of the first-order partial derivatives of $\x$, 
\ie $\Par{\jac}_{ij} = \frac{\partial\sx_i}{\partial\sxr_j}$. 
Since the mapping is polynomial, each element of $\jac$ is polynomial. 
To be well-defined, finite element formulations require the mapping between 
the reference and any physical element to be injective\cite{frey1978some}. This
imposes to the determinant of the Jacobian matrix (the Jacobian determinant) 
to be non-zero for every point of $\domref$\cite{zhang2005bijectivity} and
we conventionally impose it to be strictly positive. 
A physical element is valid if its 
Jacobian determinant is positive everywhere on the reference domain, 
otherwise it is invalid. The validity of linear simplices 
(\ie linear triangles and tetrahedra) is easy to check: since the Jacobian 
determinant is constant for these elements, it is sufficient to compute it 
at any point $\xr\in\domref$ and verify that it is positive. 
In practice, it is equivalent to compute the signed area of linear triangle since it is 
equal to the Jacobian determinant divided by 2. Similarly, the signed volume of linear
tetrahedra is equal to the Jacobian determinant divided by 6 and can equivalently be 
computed to check their validity. 
The Jacobian determinant of linear quadrangles and hexahedra, on the
other hand, 
is not constant over their reference domain. It is necessary to compute the 
minimum of their Jacobian determinant in order to check their validity. 
The two following sections are dedicated to explaining how to achieve it.

\subsection{Validity control of a linear quadrangle}

In finite element codes, the domain of definition $\domref$ of the quadrangular 
element is taken as the domain $[-1,1]\times[-1,1]$ due to better numerical properties. 
This choice has no impact on the validity criterion and we will 
consider $\domref\equiv[0,1]\times[0,1]$ in this paper for clarity reasons. 
Consequently, the Lagrange shape functions for a linear quadrangle reads:
\[
\left\{
\begin{array}{lcc}
\L{}1(\xra, \xrb) = &(1-\xra)&(1-\xrb)\\
\L{}2(\xra, \xrb) = &\xra&(1-\xrb)\\
\L{}3(\xra, \xrb) = &\xra&\xrb\\
\L{}4(\xra, \xrb) = &(1-\xra)&\phantom.\xrb.
\end{array}
\right.
\]
This implies that the mapping of a quadrangle (cf. equation~\eqref{e:mapping}), 
is bilinear. Let $(x_k, y_k)$ 
denotes the coordinates of the node $\n k$, and let us write shortly any difference 
$(x_j-x_i)$ as $\xx ij$ (and similarly for the $y$ coordinate). The partial derivative
of $x$ with respect to $\xra$ is noted $\da x$. The Jacobian matrix is given by:
\[
\jac(\xra, \xrb)=\Par{
\begin{array}{cc}
\da x & \db x\\
\da y & \db y
\end{array}
}=\Par{
\begin{array}{c@{\ \ \ }c}
\xx12\,(1-\xrb)+\xx43\,\xrb & \xx14\,(1-\xra)+\xx23\,\xra\\
\yy12\,(1-\xrb)+\yy43\,\xrb & \yy14\,(1-\xra)+\yy23\,\xra
\end{array}
}
\]
and the Jacobian determinant is given by:
\begin{align}\label{e:determinant2D}
\begin{split}
	\detJ(\xra, \xrb) = \det\jac = &\phantom{{}+{}}\L{}1(\xra, \xrb)\,\Cro{\xx12\,\yy14-\yy12\,\xx14}
				   +\L{}2(\xra, \xrb)\,\Cro{\xx12\,\yy23-\yy12\,\xx23}\\
				   &+\L{}3(\xra, \xrb)\,\Cro{\xx43\,\yy14-\yy43\,\xx14}
				   +\L{}4(\xra, \xrb)\,\Cro{\xx43\,\yy23-\yy43\,\xx23}\\
				= &\phantom{{}+{}}\sum_{k=1}^4\L{}k(\xra, \xrb)\,\detJ_k
\end{split}
\end{align}
where the coefficient $\detJ_k$ is the value taken by the Jacobian determinant 
at corner $k$. As a consequence, the Jacobian determinant is also bilinear and its minimum
is reached at one of the four corners. The validity control of linear quadrangle thus 
consists in computing the Jacobian determinant at each corner and in verifying that none 
is negative. An equivalent, but computationally more expensive test would be 
to compute the angles of the four corners and to check if they lie between $0\degree$
and $180\degree$.

\begin{figure}[ht!]
\begin{center}
\setlength\figwidth{.4\textwidth}
\input{figures/tikz/areas.tex}
\caption{The linear relationship between the areas of the triangles in a quadrangle 
($A_1+A_3=A_2+A_4$) implies an equivalent linear relationship between the four 
coefficients of the Jacobian determinant $\detJ_k$: $\detJ_1+\detJ_3 = \detJ_2+\detJ_4$, 
where $\detJ_k$ is the value taken by the Jacobian determinant 
at corner $k$.}
\label{f:areas}
\end{center}
\end{figure}
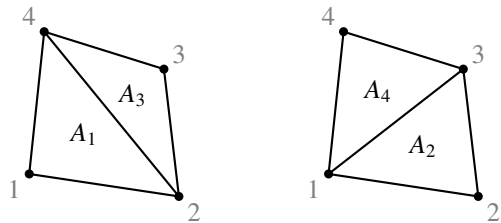

The four quantities to compute (either the angles or the coefficients $\detJ_k$) are not 
linearly independent. Indeed, concerning the angles, the existing linear relation is 
that the four angles of a quadrangle sum up to $360\degree$. Now, from
equation~\eqref{e:determinant2D}, we can deduce that the Jacobian 
determinant at \eg the first corner is equal to the third component of the vector 
$\v12\times\v14$, where $\v ij=\n j-\n i=(\xx ij, \yy ij)$ is the vector that goes 
from node $i$ to node $j$. But, for two vectors $\vec a$ 
and $\vec b$ of the $xy$-plane, it is well-known that the value of the third component 
of their cross product $\vec a\times\vec b$ is equal to the signed area of the 
parallelogram they span. In consequence, the Jacobian determinant at corner 1 is equal 
to two times the signed area of the triangle defined by $\n 1$, $\n 2$ and $\n 4$. Let us 
note $A_k$ the signed area of the triangle of corner $k$. Since the total area of the 
quadrangle is equal to $A_1+A_3$ or $A_2+A_4$, we have the following relation 
concerning the Jacobian determinant: $\detJ_1+\detJ_3 = \detJ_2+\detJ_4$ 
(see Figure~\ref{f:areas}).

\subsection{Validity control of a linear hexahedron}

Let $(x_1(\xr),\,x_2(\xr),\,x_3(\xr))$ be the trilinear mapping of the hexahedron. 
The 3D Jacobian determinant is by definition:
\beqn{e:det3d}
	\detJ(\xra,\xrb,\xrc) = \sum_{i,\,j,\,k\,=1}^3 \varepsilon_{i,\,j,\,k}
	\ \ \da{\big(x_i\big)}\ \db{\big(x_j\big)}\ \dc{\big(x_k\big)}
\eeqn
where $\varepsilon_{i,\,j,\,k}$ is the permutation symbol. We have that $\Da{x_i}$ is a 
bilinear function in $\xrb$ and $\xrc$, and similarly for $\Db{x_i}$ and $\Dc{x_i}$. 
This means that each term of the sum in equation~\eqref{e:det3d} is triquadratic and so 
is the Jacobian determinant of the linear hexahedron. As a 
consequence, the minimum of the Jacobian determinant is not 
necessarily located at one of the eight corners. A more sophisticated validity test for 
hexahedra would be to compute the minimum of $\detJ$ on the edges. This can be easily 
implemented since the Jacobian determinant restricted to an edge is a quadratic function in 
one of the reference variables. However, it has been proved in\cite{knupp1990invertibility} 
that this test is not sufficient. One step further would be the ``face test'' that would
consist in computing the global minimum of a biquadratic function 
(defined on a square domain) for the 6 faces of the hexahedron. However, there is, 
to the best of our knowledge, no proof that it would be sufficient, 
\ie that the global minimum cannot be exclusively located in the volume.

Currently, the only existing technique to robustly compute the validity of linear 
hexahedra is the method proposed in\cite{johnen2013geometrical}. This method 
computes bounds on the minimum of the Jacobian determinant that can be sharpened as 
much as desired. The main drawback of the proposed algorithm is the general framework 
used for curved elements that is not well-adapted for an efficient computation for 
the linear hexahedron. We thus propose to adapt this method to the particular case 
that concerns us.

In the next section, we introduce the B\'ezier formulation that allows to compute the 
bounds and subsequently accurately compute the minimum of $\detJ$.

\section{B\'ezier expansion of hexahedra Jacobian determinant}\label{s:bezier}

Polynomial quantities can be expanded into the so-called B\'ezier basis in order to make 
use of B\'ezier expansion properties. In this section, we first 
introduce the B\'ezier expansion, then we derive the transformation matrix that 
computes the B\'ezier coefficients from the Lagrange coefficients.

\subsection{Definition of B\'ezier expansion}

Let $\B\o k$ be the Bernstein polynomial function whose expression is:
\[
	\B\o k(t) = \nchoosek\o k\ t^k\,(1-t)^{\o-k}\qquad t\in[0,1],\ \ k=0,\dots,n
\]
where $\nchoosek\o k=\frac{\o!}{k!(\o-k)!}$ is the binomial coefficient. These functions 
allow to construct the hexahedral B\'ezier functions in term of the tensor product 
of three Bernstein polynomials:
\beqn{e:bezDefinition}
	\B\o{ijk}(\xra,\xrb,\xrc) = \B\o i(\xra)\ \B\o j(\xrb)\ \B\o k(\xrc).
\eeqn
These functions, $\{\B\o{ijk}\}_{(0\leq i,j,k\leq\o)}$, defines the 
\emph{B\'ezier basis} of the hexahedral polynomial space of order $\o$. Since the Jacobian 
determinant of the linear hexahedron is a triquadratic function, it is included 
in the hexahedral polynomial space of order $2$ and it can be expanded into 
the B\'ezier basis of order 2. There exists thus a unique set of coefficients $b_{ijk}$ 
(also known as control values) such that we have:
\beqn{e:bezExpansion}
	\detJ(\xr) = \sum_{i,j,k=0}^2\ b_{ijk}\,\B2{ijk}(\xr)
\eeqn
where the right member of the above expression is the \emph{B\'ezier expansion} of the 
Jacobian determinant. The number of coefficients is $27$ since every index can take three 
values.

B\'ezier bases have the property that the basis functions are positive over 
their domain of definition and sum up to~1. 
This implies the well-known convex hull property which, in our case, gives 
that $\min_{ijk}b_{ijk}\leq\min_{\xr}\detJ$. In addition to that, some 
B\'ezier coefficients are actual values of the Jacobian determinant. Those are the one 
``located'' at the corners of the element. For example, we have: $b_{000}=\detJ(0,0,0)$ and 
$b_{200}=\detJ(1,0,0)$. 
The minimum of these corner coefficients constitutes an upper bound for 
$\min_{\xr}\detJ$.
In other words, the control values allow to bound the minimum of the Jacobian 
determinant from below and above.
A positive lower bound implies the positivity of the Jacobian determinant and the
validity of the element. On the other hand, a negative upper bound implies that the
element is invalid. In the third and last case, when the lower bound is negative and the 
upper bound is positive, nothing can be told concerning the validity of the element.
Those bounds are not necessarily sharp. However, they can 
be sharpened as much as desired by ``subdividing'', \ie by expanding the same 
function defined on a smaller domain, called a subdomain\cite{johnen2013geometrical}. 
It is proven in\cite{leroy2008certificats,leroy2011certificats} that such 
subdivision algorithm always stops and 
that it can be used to check the positivity of a multivariate polynomials.
Moreover, the bounds converge quadratically with the size of the 
subdomains\cite{cohen1985rates}. The subdivision algorithm can be implemented in a 
recursive and adaptive manner making the validity check very 
efficient\cite{johnen2013geometrical}.

In the following section, we explain how to compute the 27 coefficients $b_{ijk}$ of the B\'ezier expansion~\eqref{e:bezExpansion}.

\subsection{Computation of the B\'ezier coefficients}

\begin{figure}[ht!]
\begin{center}
\setlength\figwidth{.3\textwidth}
\input{figures/tikz/ordering3.tex}
\caption{Ordering of the nodes. Low order nodes are in black while high order nodes are in 
gray.}
\label{f:ordercoeff}
\end{center}
\end{figure}
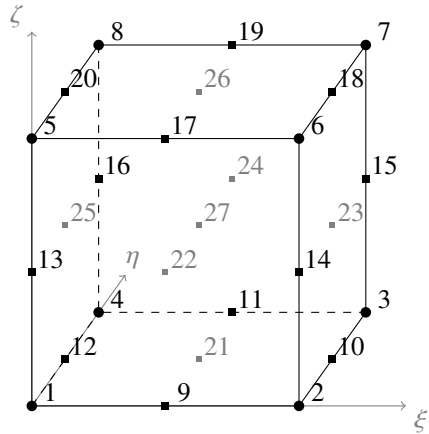

In order to compute the 27 B\'ezier coefficients we have to write a linear 
system of equations.
Let us consider a different indexing for B\'ezier coefficients and B\'ezier 
functions for which the order is given in Figure~\ref{f:ordercoeff}. This permits 
to gather the $27$ B\'ezier coefficients into a vector $\vec b$ for which we 
have, for example, $b_1\!=\!b_{000}$, $b_2\!=\!b_{200}$ and $b_9\!=\!b_{100}$.
We will use a greek letter to refer to this new indexing.
In the same way, $B_\i$ will refer to a certain function $B_{ijk}$ such 
that to respect the order defined in Figure~\ref{f:ordercoeff}. 
Let $\xr_\i, \i=1,\dots,27$ be different points of the reference domain. 
In practice, these points are taken as the uniformly spaced nodes of the order 2 
hexahedron, which limits numerical errors.
We order them in the same way, such that we have $\xr_1=(0,0,0)$, $\xr_2=(1,0,0)$ and $
\xr_9=(1/2,0,0)$ for example. Let $\vec c$ be the vector of the Jacobian determinant 
computed at those points, \ie $c_\i = \detJ(\xr_\i)$.

From the definition of the 
B\'ezier expansion~\eqref{e:bezExpansion}, we can write the following linear system:
\begin{align*}
	\detJ(\xr_\i) &= \sum_{\j=1}^{27} b_\j\,B_\j(\xr_\i)\qquad\forall \i\in\{1,\dots,27\}\\
	\Leftrightarrow\qquad \vec c &= \vec A \vec b
\end{align*}
where $\vec A$ is a transformation matrix\footnote{
Indeed, we can expand the Jacobian determinant into the traditional Lagrange 
functions of order 2 for the hexahedral element, in which case we have: $
\detJ(\xr)=\sum_{j=1}^{27}c_j\,L_j(\xr)$. The sets $\{L_j\}$ and $\{B_j\}$ are two 
different bases of the same functional space for which $\vec c$ and $\vec b$ are the 
respective coefficients of the Jacobian determinant.} in which each element 
$A_{\i\j}$ is equal to $B_\j(\xr_\i)$. The inverse of $\vec A$, denoted $\Tb$, 
is the matrix that computes the B\'ezier coefficients from the computed values 
of the Jacobian determinant, \ie $\vec b =\Tb\,\vec c$. 
Matrix $\Tb$ is given in Table~\ref{t:Tb}. 
To calculate the vector $\vec c$, one may derive 
the analytical expression of the Jacobian determinant, as we did in 2D (see equation~
\eqref{e:determinant2D}). But we will see in Section~\ref{s:tet2lag} that it can 
be performed by computing the volume of tetrahedra. Moreover, we show in the next 
section that only a small part of $\vec c$ has to be computed.
\begin{table}
\[
\Tb = \Par{\begin{array}{cccc:cccc|cccc:cccc:cccc|cccccc|c}
\multicolumn{8}{c|}{\phantom{\Big(}\identity88\phantom{\Big(}} & \multicolumn{19}{c}{\zeros8{19}}\\\hline
-\mf12&-\mf12&0&0&0&0&0&0&\multicolumn{12}{c|}{}\\
0&-\mf12&-\mf12&0&0&0&0&0&\multicolumn{12}{c|}{}\\
0&0&-\mf12&-\mf12&0&0&0&0&\multicolumn{12}{c|}{}\\
-\mf12&0&0&-\mf12&0&0&0&0&\multicolumn{12}{c|}{}\\\cdashline{1-8}[4pt/4pt]
-\mf12&0&0&0&-\mf12&0&0&0&\multicolumn{12}{c|}{}\\
0&-\mf12&0&0&0&-\mf12&0&0&\multicolumn{12}{c|}{\multirow{2}{*}{$2\,\identity{12}{12}$}}&\multicolumn{7}{c}{\multirow{2}{*}{$\zeros{12}7$}}\\
0&0&-\mf12&0&0&0&-\mf12&0&\multicolumn{12}{c|}{}\\
0&0&0&-\mf12&0&0&0&-\mf12&\multicolumn{12}{c|}{}\\\cdashline{1-8}[4pt/4pt]
0&0&0&0&-\mf12&-\mf12&0&0&\multicolumn{12}{c|}{}\\
0&0&0&0&0&-\mf12&-\mf12&0&\multicolumn{12}{c|}{}\\
0&0&0&0&0&0&-\mf12&-\mf12&\multicolumn{12}{c|}{}\\
0&0&0&0&-\mf12&0&0&-\mf12&\multicolumn{12}{c|}{}\\\hline
\mf14&\mf14&\mf14&\mf14&0&0&0&0&-1&-1&-1&-1&0&0&0&0&0&0&0&0&&&&&&&0\\\cdashline{1-20}[4pt/4pt]
\mf14&\mf14&0&0&\mf14&\mf14&0&0&-1&0&0&0&-1&-1&0&0&-1&0&0&0&&&&&&&0\\
0&\mf14&\mf14&0&0&\mf14&\mf14&0&0&-1&0&0&0&-1&-1&0&0&-1&0&0&\multicolumn{6}{c|}{\multirow{2}{*}{$4\,\identity66$}}&0\\
0&0&\mf14&\mf14&0&0&\mf14&\mf14&0&0&-1&0&0&0&-1&-1&0&0&-1&0&\multicolumn{6}{c|}{}&0\\
\mf14&0&0&\mf14&\mf14&0&0&\mf14&0&0&0&-1&-1&0&0&-1&0&0&0&-1&&&&&&&0\\\cdashline{1-20}[4pt/4pt]
0&0&0&0&\mf14&\mf14&\mf14&\mf14&0&0&0&0&0&0&0&0&-1&-1&-1&-1&&&&&&&0\\\hline
\multicolumn{8}{c|}{\phantom{\Big(}-\mf18\,\ones18\phantom{\Big(}}&\multicolumn{12}{c|}{\mf12\,\ones1{12}}&\multicolumn{6}{c|}{~-2\,\ones16~}&8
\end{array}}
\]
\caption{Transformation matrix that computes the B\'ezier coefficients from the 
sampling of the Jacobian determinant. Submatrix $\identity mm$ designate the identity 
matrix of dimension $m$, submatrix $\zeros mn$ is a $m$ by $n$ matrix with only $0$ and $
\ones mn$ is a $m$ by $n$ matrix containing only $1$.}
\label{t:Tb}
\end{table}

\section{Linear dependency of the coefficients}\label{s:only20}

Like for quadrangles, B\'ezier coefficients of the hexahedral elements are not all 
linearly independent. This is linked to the fact that the Taylor series expansion of the 
Jacobian determinant contains only $20$ non-zero coefficients, as demonstrated 
in\cite{knupp1990invertibility}. In this section, we formulate the dependency between the 
coefficients through a similar reasoning. We then construct a transformation matrix between 
the 20 linearly independent Jacobian determinant values and the $27$ B\'ezier coefficients.

The Jacobian determinant can be written as the triple scalar product:
\[
	\detJ=\triple{\da\x}{\db\x}{\dc\x}.
\]
This permits to compute the derivatives of the Jacobian determinant in terms of 
derivatives of the mapping. Given that the mapping is trilinear, the only 
non-zero derivatives of $\x$ are $\da\x$, $\db\x$, $\dc\x$, $\d\x{\xra\xrb}$, 
$\d\x{\xra\xrc}$, $\d\x{\xrb\xrc}$ and $\d\x{\xra\xrb\xrc}$. 
The derivatives of $J$ can be found in\cite{knupp1990invertibility} 
and result in the following observation:
\begin{observation}\label{p:derivative}
The following non-trivial high-order derivatives of the Jacobian determinant 
are equal to zero: $\d\detJ{\xra\xra\xrb\xrb} = \d\detJ{\xra\xra\xrc\xrc}
 = \d\detJ{\xrb\xrb\xrc\xrc} = 0$.
\end{observation}

Let us consider the monomial basis $\{M_{ijk}\}_{(0\leq i,j,k\leq2)}$, where $M_{ijk}
=M_{ijk}(\xra,\xrb,\xrc)=\xra^i\xrb^j\xrc^k$ and let us expand the Jacobian 
determinant into this basis. 
Let $m_{ijk}$ be the coefficients of this expansion. 
Observation~\ref{p:derivative} admits the following corollary:
\begin{corollary}\label{p:monomial}
7 monomial coefficients of the Jacobian determinant are always equal to zero: 
$m_{220}=m_{202}=m_{022}=m_{221}=m_{212}=m_{122}=m_{222}=0$.
\end{corollary}

Corollary~\ref{p:monomial} implies that the Jacobian determinant 
space is of dimension 20 and that it is possible to obtain 7 linear relations 
between the 27 B\'ezier/Lagrange coefficients. We will obtain them by writing the 
expression of the monomimial coefficients in function of the B\'ezier coefficients.
Let $a_{\i\j}$ be the coefficient of monomial $\i$ in the expression of the 
B\'ezier function $\j$ (whose definition is given at equation~\eqref{e:bezDefinition}). 
Mathematically, we have $B_\j(\xr) = \sum_{\i=1}^{27}a_{\i\j}\,M_\i(\xr)$. 
We can thus write:
\[
	\detJ(\xr) = \sum_{\j=1}^{27} b_\j\,B_\j(\xr) = \sum_{\i=1}^{27}\underbrace{\Cro{\sum_{\j=1}^{27}b_\j\,a_{\i\j}}}_{\displaystyle m_\i}M_\i(\xr)
\]
The linear relations between the B\'ezier coefficients are found by considering 
the equations $m_\i=\sum_{\j=1}^{27}a_{\i\j}\,b_\j$ for the 7 monomial 
coefficients of Corollary~\ref{p:monomial}. 
This leads to the matrix given in Table~\ref{t:7LinRel} that 
computes the last 7 B\'ezier coefficients in function of the first ones.
\begin{table}
\[
\vec b_{21\rightarrow27} = \Par{\begin{array}{cccc:cccc|cccc:cccc:cccc}
\mf14&\mf14&\mf14&\mf14&0&0&0&0&-\mf12&-\mf12&-\mf12&-\mf12&0&0&0&0&0&0&0&0\\\cdashline{1-20}[4pt/4pt]
\mf14&\mf14&0&0&\mf14&\mf14&0&0&-\mf12&0&0&0&-\mf12&-\mf12&0&0&-\mf12&0&0&0\\
0&\mf14&\mf14&0&0&\mf14&\mf14&0&0&-\mf12&0&0&0&-\mf12&-\mf12&0&0&-\mf12&0&0\\
0&0&\mf14&\mf14&0&0&\mf14&\mf14&0&0&-\mf12&0&0&0&-\mf12&-\mf12&0&0&-\mf12&0\\
\mf14&0&0&\mf14&\mf14&0&0&\mf14&0&0&0&-\mf12&-\mf12&0&0&-\mf12&0&0&0&-\mf12\\\cdashline{1-20}[4pt/4pt]
0&0&0&0&\mf14&\mf14&\mf14&\mf14&0&0&0&0&0&0&0&0&-\mf12&-\mf12&-\mf12&-\mf12\\\cline{1-20}
\multicolumn{8}{c|}{\phantom{\Big(}\mf14\,\ones18\phantom{\Big(}}&\multicolumn{12}{c}{-\mf14\,\ones1{12}}
\end{array}}\cdot\vec b_{20}
\]
\caption{Computation of the last 7 B\'ezier coefficients in function of the 20 first.
}
\label{t:7LinRel}
\end{table}
Let us write $\vec D$ the matrix that computes the 27 B\'ezier coefficients from 
the first 20 B\'ezier coefficients. Matrix $\vec D$ is constructed by extending 
the matrix given in Table~\ref{t:7LinRel} with an identity matrix of size 20.
We have:
\[
  \vec b = \vec D\,\vec b_{20}
\]
where $\vec b_{20}$ is the vector containing the first 20 components of $\vec b$.

Constructing the matrix that computes the 27 B\'ezier coefficients in function of 
20 Lagrange coefficients is now straightforward. Matrix $\Tb$ (see Table~\ref{t:Tb}) 
is such that the first 20 B\'ezier coefficients depends only on the first 20 Lagrange 
coefficients. Let $\Tsub$ be the $20\times20$ upper left submatrix of $\Tb$ and $\vec 
c_{20}$ the first 20 components of $\vec c$. We have that:
\[
	\vec b_{20}=\Tsub\,\vec c_{20}\quad\Leftrightarrow\quad\vec b = 
	\vec D\,\Tsub\vec c_{20} = \Tsmall\vec c_{20}
\]
where $\Tsmall$, the matrix that computes all the B\'ezier coefficients from 
the first 20 Lagrange coefficients, is given in Table~\ref{t:Tsmall}.
\begin{table}
\[
\Tsmall = \Par{\begin{array}{cccc:cccc|cccc:cccc:cccc}
\multicolumn{8}{c|}{\phantom{\Big(}\identity88\phantom{\Big(}} & \multicolumn{12}{c}{\zeros8{12}}\\\hline
-\mf12&-\mf12&0&0&0&0&0&0&\\
0&-\mf12&-\mf12&0&0&0&0&0&\\
0&0&-\mf12&-\mf12&0&0&0&0&\\
-\mf12&0&0&-\mf12&0&0&0&0&\\\cdashline{1-8}[4pt/4pt]
-\mf12&0&0&0&-\mf12&0&0&0&\\
0&-\mf12&0&0&0&-\mf12&0&0&\multicolumn{12}{c}{\multirow{2}{*}{$2\,\identity{12}{12}$}}\\
0&0&-\mf12&0&0&0&-\mf12&0&\\
0&0&0&-\mf12&0&0&0&-\mf12&\\\cdashline{1-8}[4pt/4pt]
0&0&0&0&-\mf12&-\mf12&0&0&\\
0&0&0&0&0&-\mf12&-\mf12&0&\\
0&0&0&0&0&0&-\mf12&-\mf12&\\
0&0&0&0&-\mf12&0&0&-\mf12&\\\hline
-\mf34&-\mf34&-\mf34&-\mf34&0&0&0&0&~1~&~1~&~1~&~1~&~0~&~0~&~0~&~0~&~0~&~0~&~0~&~0~\\\cdashline{1-20}[4pt/4pt]
-\mf34&-\mf34&0&0&-\mf34&-\mf34&0&0&1&0&0&0&1&1&0&0&1&0&0&0\\
0&-\mf34&-\mf34&0&0&-\mf34&-\mf34&0&0&1&0&0&0&1&1&0&0&1&0&0\\
0&0&-\mf34&-\mf34&0&0&-\mf34&-\mf34&0&0&1&0&0&0&1&1&0&0&1&0\\
-\mf34&0&0&-\mf34&-\mf34&0&0&-\mf34&0&0&0&1&1&0&0&1&0&0&0&1\\\cdashline{1-20}[4pt/4pt]
0&0&0&0&-\mf34&-\mf34&-\mf34&-\mf34&0&0&0&0&0&0&0&0&1&1&1&1\\\hline
\multicolumn{8}{c|}{\phantom{\Big(}-\mf58\,\ones18\phantom{\Big(}}&\multicolumn{12}{c}{\mf12\,\ones1{12}}
\end{array}}
\]
\caption{Transformation matrix that allows to compute the B\'ezier coefficients from 20 
samplings of the Jacobian determinant. 
matrix of dimension $m$, submatrix $\zeros mn$ is a $m$ by $n$ matrix with only $0$ and $
\ones mn$ is a $m$ by $n$ matrix containing only $1$.
}
\label{t:Tsmall}
\end{table}

\section{Expression of the 20 Lagrange coefficients in function of 20 tetrahedral volumes}
\label{s:tet2lag}

In this section we show that the 20 Lagrange coefficients that has to be computed are
equal to the volume of tetrahedra.

Recalling that $\x$ is the column vector $(x, y, z)^{\text T}$, the Jacobian matrix can be 
written as:
\[
\jac(\xra,\xrb,\xrc)=\Par{
\,\Cro{\begin{array}{c}\\\da\x\\~\end{array}}\,
\Cro{\begin{array}{c}\\\db\x\\~\end{array}}\,
\Cro{\begin{array}{c}\\\dc\x\\~\end{array}}\,
},
\]
Let us recall that $\v\i\j$ denotes the difference $(\n\j-\n\i)$. 
We can express the derivatives of $\x$ from the definition of the 
mapping~\eqref{e:mapping} and the Lagrange functions given in~\ref{s:lagrangeFunction3D}:
\[
\left\{
\begin{array}{l}
\da\x = \v12\ (1-\xrb)\,(1-\xrc) + \v43\ \xrb\,(1-\xrc) + \v56\ (1-\xrb)\,\xrc + \v87\ \xrb\,\xrc\\
\db\x = \v14\ (1-\xra)\,(1-\xrc) + \v23\ \xra\,(1-\xrc) + \v58\ (1-\xra)\,\xrc + \v67\ \xra\,\xrc\\
\dc\x = \v15\ (1-\xra)\,(1-\xrb) + \v26\ \xra\,(1-\xrb) + \v48\ (1-\xra)\,\xrb + \v37\ \xra\,\xrb.
\end{array}
\right.
\]

In the following, $\det{\vec a, \vec b, \vec c}$ will denote the determinant of the matrix 
made up of columns $\vec a$, $\vec b$ and $\vec c$. Note that $\det{\vec a, \vec b, \vec c}
$ equals $\triple{\vec a}{\vec b}{\vec c}$ and is a trilinear function. Moreover, if the 
three vectors have the same origin, then the determinant is also 6 times the volume of the 
tetrahedron that the vectors define. Lastly, if the three vectors are not 
linearly independent, then the determinant is zero.

There are two types of Lagrange coefficients we are interested in: the coefficients that 
correspond to the corners and the coefficients that correspond to the edges of the 
hexahedron. By symmetry of the problem, there must be also two types of tetrahedra to 
identify. It is already well-known that the Jacobian determinant computed at a corner 
corresponds to 6 times the volume of the tetrahedron constructed from the 3 edges of the 
corner. Let us formulate it mathematically for the first corner:
\[
\detJ_1 = \det{\jac(0,0,0)} = \det{\v12,\,\v14,\,\v15} = 6\ \text{vol}\,(\n1,\,\n2,\,\n4,\,\n5)
\]
where $\vol(\cdot)$ refer to the volume of the tetrahedron defined by the four nodes.

In a similar manner, we can express the 9th value of the Jacobian determinant as the volume 
of a tetrahedron:
\begin{align*}
	\detJ_9 = \det{\jac(1/2,0,0)} &= \det{\v12,\ \frac{\v14+\v23}2,\ \frac{\v15+\v26}2} \\
	&= \det{\v12,\ \frac{\n4+\n3}2-\frac{\n1+\n2}2,\ \frac{\n5+\n6}2-\frac{\n1+\n2}2} \\
	&= \det{\v12,\ \Cro{\frac{\n4+\n3}2-\n1}+\Cro{\n1-\frac{\n1+\n2}2},\ \Cro{\frac{\n5+\n6}2-\n1}+\Cro{\n1-\frac{\n1+\n2}2}} 
\end{align*}
where the terms $\Cro{\n1-\frac{\n1+\n2}2}$ are equal to $-\frac{\v12}2$. By trilinearity 
of the determinant and dependency with respect to the first vector ($\v12$), the terms 
$-\frac{\v12}2$ vanish and we obtain:
\[
	\detJ_9 = 6\ \text{vol}\,\Par{\n1,\ \n2,\ \frac{\n4+\n3}2,\ \frac{\n5+\n6}2}
\]

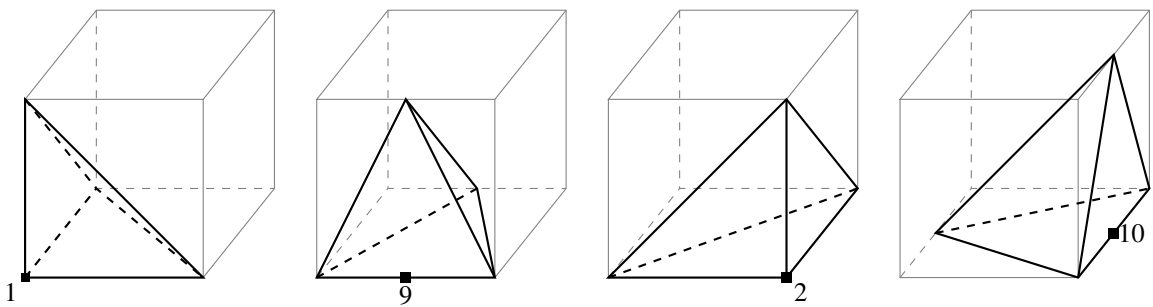
\begin{figure}[ht!]
\begin{center}
\setlength\figwidth{.2\textwidth}
\input{figures/tikz/tetrahedron1.tex}
\input{figures/tikz/tetrahedron9.tex}
\input{figures/tikz/tetrahedron2.tex}
\input{figures/tikz/tetrahedron10.tex}
\caption{Different tetrahedra whose volume corresponds to the value of the respective 
coefficients $\detJ_1$, $\detJ_9$, $\detJ_2$ and $\detJ_{10}$ divided by $6$.}
\label{f:tet}
\end{center}
\end{figure}

Figure~\ref{f:tet} shows the tetrahedra that correspond to four value of the Jacobian 
determinant.

\section{The algorithm}\label{s:algo}

The algorithm that computes the validity of a linear hexahedron takes as input 
the 8 nodes coordinates of the element. 
It returns true if the element is valid and return false if the element is invalid. 
The execution is the following:
\begin{enumerate}
\item Compute the 20 volumes of Section~\ref{s:tet2lag} and put them in vector $\vec v$
(ordering them as in Figure~\ref{f:ordercoeff}).
\item If at least one volume is negative, return False.
\item Compute the B\'ezier coefficients $\vec b = \Tsmall\vec v$ where $\Tsmall$ 
is the matrix given in Table~\ref{t:Tsmall}.
\item If all the B\'ezier coefficients in $\vec b_{9\to27}$ are positive, return True.
\item Return \texttt{recursive\_subdivision}($\vec b$).
\end{enumerate}
In Step~4, the 8 first B\'ezier coefficients are equal to the volume of the corner
tetrahedra and must be positive otherwise the algorithm would have stop at Step~2.

The subdivision algorithm, \texttt{recursive\_subdivision}($\vec b$), is identical 
to the subdivision algorithm presented in paper\cite{johnen2013geometrical}
(although implemented in a more efficient manner in our new implementation). 
It takes a vector of 27 B\'ezier coefficients as input and return true if the 
Jacobian determinant is strictly positive on the subdomain, otherwise it returns
false. The algorithm is:
\begin{enumerate}
\item Subdivide: Compute the subcoefficients $\vec b^i,\ i=1,\dots,8$ as described 
in paper\cite{johnen2013geometrical}.
\item For each $\vec b^i$:
\item \ \ If at least one of the coefficients in $\vec b^i_{1\to8}$ is negative, 
return False.
\item \ \ If all the coefficients in $\vec b^i_{9\to27}$ are positive, continue the loop.
\item \ \ If \texttt{recursive\_subdivision}($\vec b^i$) is false, return False.
\item Return True.
\end{enumerate}
In Step~3 of this algorithm, it is checked if the 8 first B\'ezier coefficients are 
not negative since they are actual values of the Jacobian determinant. In Step~4, 
the positivity of the 19 other coefficients ensures that the Jacobian determinant is 
positive on the corresponding subdomain in which case the algorithm skip Step~5 and
continue the loop. While there is no negative real value of the Jacobian 
determinant but at least one negative B\'ezier coefficients, the algorithm 
subdivide (Step~5).

\section{Results}\label{s:results}

\begin{figure}[htp]
\begin{center}
\begin{tikzpicture}
\node[anchor=east] at (0, 0) {
\begin{tabular}{c|ccc}
i & x & y & z\\
\hline
0 & 0 &  0 & 0 \\
1 & 1 & 0 & 0 \\
2 & 1.7615170641459 & 0.594764272968121 & 0.15552188663289 \\
3 & 0.438888411629833 & 1.53098020041072 & 0.185631029838277 \\
4 & 1.3859049651391 & 0.0755794018509022 & 1.77483024073906 \\
5 & 1.22129676447071 & 0.271876165350328 & 0.630922158503566 \\
6 & 1.77365642274365 & 1.25103990471942 & 1.83300604452892 \\
7 & 0.0769922201302364 & 0.940424880836765 & 1.45521546591891 \\
\end{tabular}
};
\node[anchor=west] at (0, 0) {
\includegraphics[width=.3\textwidth]{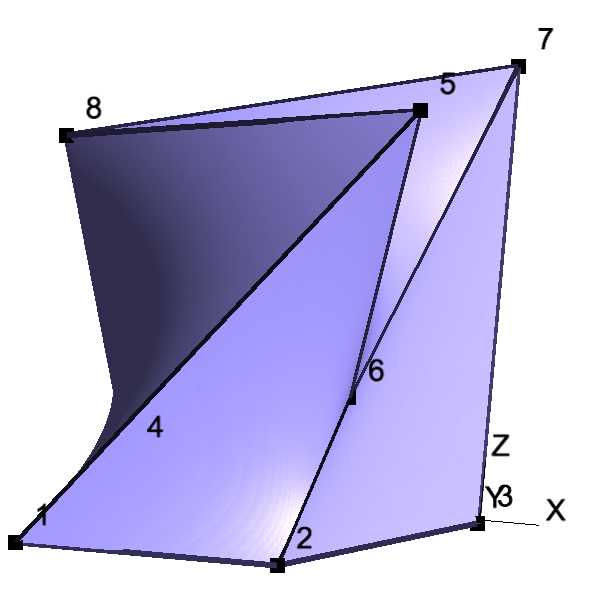}
};
\end{tikzpicture}
\end{center}
\caption{Invalid hexahedron for which the Jacobian determinant is positive at the
$27$ nodes of the second-order hexahedron and for which the 24 tetrahedral volumes of 
Ushakova's\cite{ushakova2011nondegeneracy} test6 are all positive.}
\label{f:invalidButUshakovaValid}
\end{figure}

\begin{figure}[htp]
\begin{center}
\begin{tikzpicture}
\node[anchor=east] at (0, 0) {
\begin{tabular}{c|ccc}
i & x & y & z\\
\hline
0 & 0 &0&  0\\
1 &1 &0& 0 \\
2 &1.539& 0.704696 &1.84011\\
3 & 0.166589 &1.08208 &0.162539\\
4 & 0.0501127& 1.96347 &1.56559\\
5 & 0.422336 &0.00419138& 1.43038\\
6 & 0.509917 & 0.0214216 &1.55322\\
7 & 0.40783 & 1.73452 &  1.93234\\
\end{tabular}
};
\node[anchor=west] at (0, 0) {
\includegraphics[width=.3\textwidth]{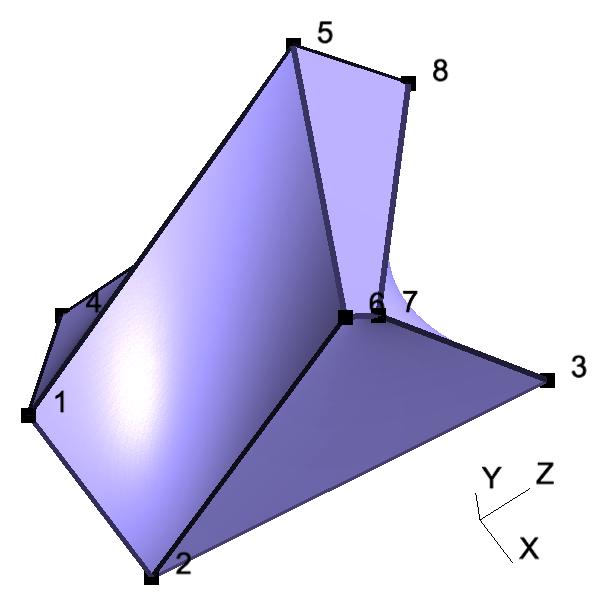}
};
\end{tikzpicture}
\end{center}
\caption{Valid hexahedron that does not pass Ushakova's\cite{ushakova2011nondegeneracy} test6.}
\label{f:validNegativeOshakova}
\end{figure}

\begin{figure}[htp]
\begin{center}
\begin{tikzpicture}
\node[anchor=east] at (0, 0) {
\begin{tabular}{c|ccc}
i & x & y & z\\
\hline
1 & 0.464949491866817 & 0.358989226966155 & 0.0133365886410108\\
2 & 0.481795709097567 & 0.358745078890347 & 0.0163884395886105\\
3 & 0.482406079287087 & 0.351664784691916 & 0.0235297708059938\\
4 & 0.466719565416425 & 0.339945677053133 & 0.0278023621326335\\
5 & 0.465498825037385 & 0.320291756950591 & -0.00277718436231578\\
6 & 0.465987121189001 & 0.321085238196966 & -0.0042420728171636\\
7 & 0.501998962370677 & 0.322367015594958 & -0.0116275521103549\\
8 & 0.487166966765343 & 0.308816797387616 & 0.0115054780724508
\end{tabular}
};
\node[anchor=west] at (0, 0) {
\includegraphics[width=.3\textwidth]{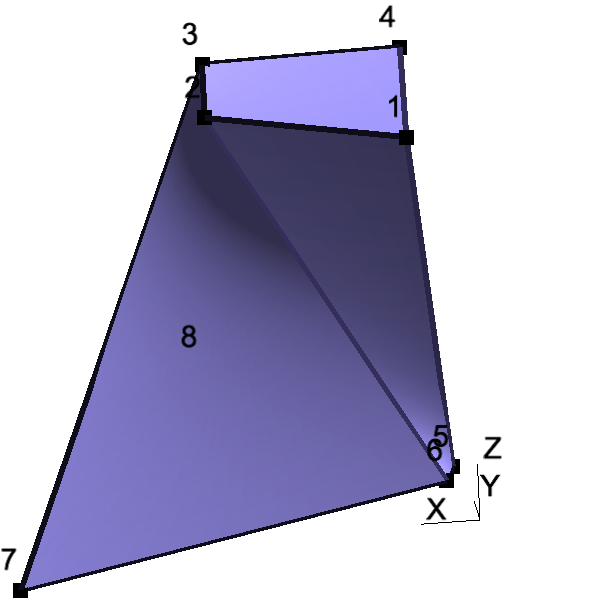}
};
\end{tikzpicture}
\end{center}
\caption{Invalid hexahedron for which the minimum of the scaled Jacobian computed at the
corners is equal to 0.64.}
\label{f:invalidHighCornerQual}
\end{figure}

We begin the results with unitary tests. The Jacobian determinant of the hexahedron 
defined in Figure~\ref{f:invalidButUshakovaValid} is positive at the 8 corners, the 
center of the edges, the center of the faces and the center of the volume. 
Moreover, the hexahedron passes the Ushakova's\cite{ushakova2011nondegeneracy} test6 
that requires the computation of $24$ tetrahedral volumes.
Our algorithm detects that this hexahedron is invalid. 
Figure~\ref{f:validNegativeOshakova} presents a hexahedron that does not pass
Ushakova's\cite{ushakova2011nondegeneracy} test6, despite the fact that the
element is valid.
In hexahedral mesh community, it is common to measure the quality of hexahedra by 
computing the minimum of the ``scaled Jacobian'' on the 8 
corners\cite{knupp2000achievingB,yamakawa2003fully}.
For the hexahedron of Figure~\ref{f:invalidHighCornerQual}, this quality measure 
is equal to 0.64 although the element is invalid. This demonstrates that even 
invalid hexahedra can have a good quality at the corners.
%
%

For the next experimentation, we compare our method with some previous methods 
on different datasets. The results are given in Table~\ref{t:comparison}.
The datasets have been generated by the algorithm 
described in\cite{pellerin2017} which takes a tetrahedral mesh as input and computes 
hexahedra that can be created by combining tetrahedra. This algorithm can generate a 
large amount of hexahedra of different qualtity. 
We have considered two models. The first one, ``Fusee\_1'',
contains $71,947$ vertices and $349,893$ tetrahedra. The second one is ``FT47'' and 
contains $370,401$ vertices and $2,085,394$ tetrahedra. Both of them are available
on the website \url{www.hextreme.eu}. 
We have disabled the validity check during the hexahedra creation and, for each model, 
we have generated three datasets of hexahedra by varying the desired minimal quality $q$
(computed at the corners). Datasets that correspond to $q=-1$ contain a large proportion
of invalid hexahedra while datasets that correspond to $q=0.5$ contain only valid 
hexahedra. We have compared our new implementation with the previous 
one\cite{johnen2013geometrical}, as well as the 5 first validity tests presented 
in\cite{ushakova2011nondegeneracy}. These tests consist in computing the volume of 
respectively 8, 10, 24, 32 and 58 tetrahedra and returning False as soon as a 
negative volume is found or returning True if no negative volume is obtained. 
For each algorithm we store the execution time as well as the 
number of false valid (the number of invalid hexahedra that pass the test) and the 
number of false invalid (the number of valid hexahedra that do not pass the test).
The experimentation has been conducted in serial on a MacBook Pro 2016 @ 2.9 GHz.

Our new implementation detects the same invalid hexahedra than our previous 
implementation. We have taken this result as the reference for computing the false 
invalid and false valid elements of the methods from\cite{ushakova2011nondegeneracy}. 
Test~1 computes the volume of corner tetrahedra, which corresponds to a necessary 
condition. As expected, Test~1 misses invalid elements but never finds false invalid.
Test~2 to Test~4 are neither sufficient nor necessary. Test~4 misses very few
invalid hexahedra, however. 
Test~5 corresponds to a sufficient condition and can miss as much as 80\% of
valid elements (see dataset Fusee\_1, $q=-1$).

Our new implementation is about $15$ to $30$ time faster than the algorithm 
designed for curvilinear elements and runs at similar speed than Test~5 
of\cite{ushakova2011nondegeneracy} which consists in computing $58$ tetrahedral 
volumes. Our new algorithm can check the validity of hexahedra at a rate of 
between $6$ million and $12$ million hexahedra per second on a single core. 
The speed is higher when there is a large proportion of invalid 
hexahedra since the algorithm can stop at an early stage if a negative 
Jacobian determinant is obtained.

\begin{table}
\renewcommand{\arraystretch}{1.}  
\setlength{\tabcolsep}{3pt} 
\centering
\scriptsize
\caption{Comparison of our new implementation with some previous methods. The datasets 
differ in the number of hexahedra and proportion of invalid element amongst them.
For each method, the computation time, the number of false valid and the number
of false invalid are given. Numbers are given with 3 significant digits.}
\begin{tabular}{lrr@{\qquad}l@{\quad}rrrrrrr}
\toprule
Dataset & \# hex & \# invalid & & Ours & \citet{johnen2013geometrical}
 & Test 1\cite{ushakova2011nondegeneracy} & Test 2\cite{ushakova2011nondegeneracy}
 & Test 3\cite{ushakova2011nondegeneracy} & Test 4\cite{ushakova2011nondegeneracy}
 & Test 5\cite{ushakova2011nondegeneracy} \\
\midrule
Fusee\_1, $q=0.5$ & $334,000$ & $0$ & \# false valid & $0$ & $0$ & $0$ & $0$ & $0$ & $0$ & $0$\\
 & & & \# false invalid & $0$ & $0$ & $0$ & $529$ & $10,000$ & $10,000$ & $111,000$\\[.5ex]
 & & & time [s] & $0.0565$ & $0.812$ & $0.0180$ & $0.0168$ & $0.0349$ & $0.0549$ & $0.0866$\\[2ex]
%
Fusee\_1, $q=0$ & $2,040,000$ & $79,900$ & \# false valid & $0$ & $0$ & $79,900$ & $73,900$ & $0$ & $0$ & $0$\\
 & & & \# false invalid & $0$ & $0$ & $0$ & $10,200$ & $814,000$ & $814,000$ & $1,590,000$\\[.5ex]
 & & & time [s] & $0.339$ & $6.60$ & $0.0945$ & $0.0977$ & $0.161$ & $0.220$ & $0.358$\\[2ex]
%
Fusee\_1, $q=-1$ & $6,060,000$ & $4,110,000$ & \# false valid & $0$ & $0$ & $80,000$ & $74,000$ & $48,400$ & $0$ & $0$\\
 & & & \# false invalid & $0$ & $0$ & $0$ & $10,200$ & $814,000$ & $814,000$ & $1,590,000$\\[.5ex]
 & & & time [s] & $0.488$ & $15.5$ & $0.202$ & $0.212$ & $0.301$ & $0.347$ & $0.418$\\[2ex]
%
FT47, $q=0.5$ & $3,000,000$ & $0$ & \# false valid & $0$ & $0$ & $0$ & $0$ & $0$ & $0$ & $0$\\
 & & & \# false invalid & $0$ & $0$ & $0$ & $1,890$ & $115,000$ & $115,000$ & $1,060,000$\\[.5ex]
 & & & time [s] & $0.459$ & $6.93$ & $0.181$ & $0.203$ & $0.341$ & $0.463$ & $0.725$\\[2ex]
%
FT47, $q=0$ & $14,700,000$ & $366,000$ & \# false valid & $0$ & $0$ & $366,000$ & $342,000$ & $7$ & $7$ & $0$\\
 & & & \# false invalid & $0$ & $0$ & $0$ & $38,900$ & $4,880,000$ & $4,880,000$ & $11,100,000$\\[.5ex]
 & & & time [s] & $2.43$ & $42.5$ & $0.712$ & $0.872$ & $1.30$ & $1.73$ & $2.33$\\[2ex]
%
FT47, $q=-1$ & $40,500,000$ & $26,100,000$ & \# false valid & $0$ & $0$ & $370,000$ & $346,000$ & $247,000$ & $7$ & $0$\\
 & & & \# false invalid & $0$ & $0$ & $0$ & $38,900$ & $4,880,000$ & $4,880,000$ & $11,100,000$\\[.5ex]
 & & & time [s] & $3.17$ & $102$ & $1.55$ & $1.54$ & $2.08$ & $2.49$ & $3.09$\\
%
\bottomrule
\end{tabular}%
\label{t:comparison}%
\end{table}%

\section{Conclusion}\label{s:conclusion}
Our implementation is able to check the validity of linear hexahedral elements 
in a very efficient manner. The algorithm benefit from the robustness of the 
previous method for checking the validity of curvilinear 
elements\cite{johnen2013geometrical} on which it is based.
The novelty consists of two improvements: (1) a reduced number of quantities
to be computed at the beginning of the algorithm and (2) the computation of those 
quantities as tetrahedral volumes instead of the Jacobian determinant. The 
particularization to hexahedra also permits a fine-tuned implementation.
Our new code runs more than 15 time faster than the previous code for curvilinear
elements and runs at similar speed than the sufficient but not necessary method 
presented in\cite{ushakova2011nondegeneracy}.
More than $6$ million hexahedra per second can be analyzed on a single core of 
a personal computer.
The algorithm is simple and can readily be implemented from the information
given in this paper. The C++ code will be available in Gmsh 
(\url{www.gmsh.info}).

\section*{Acknowledgements}

This research project was funded by the European Research Council 
(project HEXTREME, ERC-2015-AdG-694020) and the TILDA project. The TILDA (Towards 
Industrial LES/DNS in Aeronautics - Paving the Way for Future Accurate CFD) 
project has received funding from the European Union’s Horizon 2020 research 
and innovation program under grant agreement No 635962.
The project is a collaboration between NUMECA, DLR, ONERA, DASSAULT, SAFRAN, CERFACS, 
CENAERO, UCL, UNIBG, ICL and TsAGI.

\appendix

\section{Lagrange shape functions of the linear hexahedron}
\label{s:lagrangeFunction3D}

In this paper, we consider the following Lagrange shape functions for the 
linear hexahedron:
\[
\left\{
\begin{array}{lccc}
\L{}1(\xra, \xrb, \xrc) = &(1-\xra)&(1-\xrb)&(1-\xrc)\\
\L{}2(\xra, \xrb, \xrc) = &\xra&(1-\xrb)&(1-\xrc)\\
\L{}3(\xra, \xrb, \xrc) = &\xra&\xrb&(1-\xrc)\\
\L{}4(\xra, \xrb, \xrc) = &(1-\xra)&\xrb&(1-\xrc)\\
\L{}5(\xra, \xrb, \xrc) = &(1-\xra)&(1-\xrb)&\xrc\\
\L{}6(\xra, \xrb, \xrc) = &\xra&(1-\xrb)&\xrc\\
\L{}7(\xra, \xrb, \xrc) = &\xra&\xrb&\xrc\\
\L{}8(\xra, \xrb, \xrc) = &(1-\xra)&\xrb&\phantom{.}\xrc.
\end{array}
\right.
\]

\bibliographystyle{model1-num-names}
\bibliography{bib_minSampleHex,bib_minSampleHex2}

\end{document}

%% file: figures/tikz/mapping.tex
\begin{tikzpicture}
\usetikzlibrary{calc}

\makeatletter
\@ifundefined{Hpict}{\newlength\Hpict}{}
\@ifundefined{Wpict}{\newlength\Wpict}{}
\@ifundefined{width}{\newlength\width}{}
\makeatother

\setlength\Hpict{\figheight}
\setlength\Wpict{\figwidth}
\setlength\width{.35\Wpict}

\tikzstyle{line} = [black]
\tikzstyle{tline} = [thick, black]
\tikzstyle{vertex} = [circle, inner sep = 1.2pt, fill = black]

\node[coordinate](R0) at(0, 0) {};
\node[coordinate](R1) at($(R0) + (.975\width,-.05\width)$) {};
\node[coordinate](R2) at($(R0) + (.15\width,.3\width)$) {};
\node[coordinate](R3) at($(R0) + (0,.95\width)$) {};
\node[coordinate](r1) at($(R0) + .8*(R1)-.8*(R0)$) {};
\node[coordinate](r2) at($(r1) + .8*(R2)-.8*(R0)$) {};
\node[coordinate](r3) at($(R0) + .8*(R2)-.8*(R0)$) {};
\node[coordinate](r4) at($(R0) + .8*(R3)-.8*(R0)$) {};
\node[coordinate](r5) at($(r1) + .8*(R3)-.8*(R0)$) {};
\node[coordinate](r6) at($(r2) + .8*(R3)-.8*(R0)$) {};
\node[coordinate](r7) at($(r3) + .8*(R3)-.8*(R0)$) {};
\draw[gray,->] (R0)--(R1);
\draw[gray,->] (R0)--(R2);
\draw[gray,->] (R0)--(R3);
\node[gray] at($(R0)+(.5\width,.5\width)$) {$\domref$};
\node[vertex] (r) at($(R0)+(.25\width,.15\width)$) {};
\node[anchor=west] at($(r)+(0,0)$) {$\xr$};
\node[gray] at($(R1)+(.2,-.2)$) {$\xi$};
\node[gray] at($(R2)+(.2,.2)$) {$\eta$};
\node[gray] at($(R3)+(-.2,.2)$) {$\zeta$};
\draw[thick] (r4)--(R0)--(r1)--(r5)--(r4)--(r7)--(r6)--(r5)  (r1)--(r2)--(r6);
\draw[thick,dashed] (R0)--(r3)--(r7)  (r2)--(r3);

\node[coordinate](P0) at(.7\Wpict, 0) {};
\node[coordinate](P1) at($(P0) + (.975\width,-.05\width)$) {};
\node[coordinate](P2) at($(P0) + (.15\width,.3\width)$) {};
\node[coordinate](P3) at($(P0) + (0,.97\width)$) {};
\node[coordinate,shift={(.2,.2)}](p0) at($(P0)+(.15\width, .1\width)$) {};
\node[coordinate,shift={(.2,.2)}](p1) at($(P0)+(.8\width, .2\width)$) {};
\node[coordinate,shift={(.2,.2)}](p2) at($(P0)+(.9\width, .4\width)$) {};
\node[coordinate,shift={(.2,.2)}](p3) at($(P0)+(.35\width, .4\width)$) {};
\node[coordinate,shift={(.2,.2)}](p4) at($(P0)+(.2\width, .65\width)$) {};
\node[coordinate,shift={(.2,.2)}](p5) at($(P0)+(.7\width, .55\width)$) {};
\node[coordinate,shift={(.2,.2)}](p6) at($(P0)+(.8\width, .75\width)$) {};
\node[coordinate,shift={(.2,.2)}](p7) at($(P0)+(.35\width, .8\width)$) {};
\draw[gray, ->] (P0)--(P1);
\draw[gray, ->] (P0)--(P2);
\draw[gray, ->] (P0)--(P3);
\node[vertex,shift={(.2,.2)}] (p) at($(P0)+(.4\width,.3\width)$) {};
\node[anchor=west] at($(p)+(0,0)$) {$\x(\xr)$};
\node[gray] at($(P1)+(.2,-.2)$) {$x$};
\node[gray] at($(P2)+(-.2,.2)$) {$y$};
\node[gray] at($(P3)+(-.2,.2)$) {$z$};
\draw[thick] (p4)--(p0)--(p1)--(p5)--(p4)--(p7)--(p6)--(p5)  (p1)--(p2)--(p6);
\draw[thick,dashed] (p0)--(p3)--(p7)  (p2)--(p3);

\makeatletter
\@ifundefined{matJ}{%
	\def\matJ{\vec J}%
}{}
\makeatother
\node[coordinate](m0a) at($-.6*(R0)+1.1*(R1)+.5*(R3)$) {};
\node[coordinate](m0b) at($.7*(P0)+.5*(P3)-.2*(P1)$) {};
\node[coordinate](m0) at($.5*(m0a)+.5*(m0b) + .05*(R3) - .05*(R0)$) {};
\draw[thick, ->, >=stealth] plot [smooth, tension=1] coordinates {(m0a) (m0) (m0b)};
\node at($(m0)+(-.1,.3)$) {$\phi$};

\end{tikzpicture}

%% file: figures/tikz/areas.tex
\begin{tikzpicture}
\usetikzlibrary{calc}

\makeatletter
\@ifundefined{Hpict}{\newlength\Hpict}{}
\@ifundefined{Wpict}{\newlength\Wpict}{}
\@ifundefined{width}{\newlength\width}{}
\makeatother

\setlength\Hpict{\figheight}
\setlength\Wpict{\figwidth}
\setlength\width{.3\Wpict}

\tikzstyle{line} = [black]
\tikzstyle{tline} = [thick, black]
\tikzstyle{vertex} = [circle, inner sep = 1.2pt, fill = black]

\node[coordinate](1) at(0, 0) {};
\node[coordinate](2) at($(1) + (1\width,-.15\width)$) {};
\node[coordinate](3) at($(1) + (.9\width,.7\width)$) {};
\node[coordinate](4) at($(1) + (.1\width,.95\width)$) {};
\node[coordinate](dx) at(2\width, 0) {};
\node[coordinate](b1) at($(1) + (dx)$) {};
\node[coordinate](b2) at($(2) + (dx)$) {};
\node[coordinate](b3) at($(3) + (dx)$) {};
\node[coordinate](b4) at($(4) + (dx)$) {};
\draw[thick] (1)--(2)--(3)--(4)--cycle  (2)--(4);
\draw[thick] (b1)--(b2)--(b3)--(b4)--cycle  (b1)--(b3);
\node at($.333*(1)+.333*(2)+.333*(4)$) {$A_1$};
\node at($.333*(3)+.333*(2)+.333*(4)+(.05,.05)$) {$A_3$};
\node at($.333*(b2)+.333*(b1)+.333*(b3)$) {$A_2$};
\node at($.333*(b4)+.333*(b1)+.333*(b3)$) {$A_4$};
\node[vertex] at(1) {};
\node[vertex] at(2) {};
\node[vertex] at(3) {};
\node[vertex] at(4) {};
\node[vertex] at(b1) {};
\node[vertex] at(b2) {};
\node[vertex] at(b3) {};
\node[vertex] at(b4) {};
\node[gray] at($(1)+(-.2,-.2)$) {$1$};
\node[gray] at($(2)+(.2,-.2)$) {$2$};
\node[gray] at($(3)+(.2,.2)$) {$3$};
\node[gray] at($(4)+(-.2,.2)$) {$4$};
\node[gray] at($(b1)+(-.2,-.2)$) {$1$};
\node[gray] at($(b2)+(.2,-.2)$) {$2$};
\node[gray] at($(b3)+(.2,.2)$) {$3$};
\node[gray] at($(b4)+(-.2,.2)$) {$4$};

\end{tikzpicture}

%% file: figures/tikz/ordering3.tex
\begin{tikzpicture}
\usetikzlibrary{calc}

\makeatletter
\@ifundefined{width}{\newlength\width}{}
\makeatother

\setlength\width{.714\figwidth}

\tikzstyle{line} = [black]
\tikzstyle{tline} = [thick, black]
\tikzstyle{vertex} = [circle, inner sep = 1.4pt, fill = black]
\tikzstyle{vertexe} = [rectangle, inner sep = 1.4pt, fill = black]
\tikzstyle{vertexf} = [rectangle, inner sep = 1pt, fill = gray]
\tikzstyle{vertexv} = [rectangle, inner sep = 1pt, fill = gray]

\node[coordinate](third) at($(.25*\width,.35*\width)$) {};
\node[coordinate](t) at($(.07*\width,.05*\width)$) {};

\node[coordinate](a1) at(0, 0) {};
\node[coordinate](a2) at($(a1) + (\width,0)$) {};
\node[coordinate](a3) at($(a2) + (third)$) {};
\node[coordinate](a4) at($(a1) + (third)$) {};
\node[coordinate](a5) at($(a1) + (0,\width)$) {};
\node[coordinate](a6) at($(a5) + (\width,0)$) {};
\node[coordinate](a7) at($(a6) + (third)$) {};
\node[coordinate](a8) at($(a5) + (third)$) {};

\node[coordinate](r1) at($1.4*(a2)-.4*(a1)$) {};
\node[coordinate](r2) at($1.4*(a4)-.4*(a1)$) {};
\node[coordinate](r3) at($1.4*(a5)-.4*(a1)$) {};
\draw[gray,->] (a1)--(r1);
\draw[gray,->] (a1)--(r2);
\draw[gray,->] (a1)--(r3);
\node[gray] at($(r1)+(.2,-.2)$) {$\xi$};
\node[gray] at($(r2)+(.1,.2)$) {$\eta$};
\node[gray] at($(r3)+(-.2,.2)$) {$\zeta$};

\draw[] (a5)--(a1)--(a2)--(a6)--(a5)--(a8)--(a7)--(a3)--(a2);
\draw[] (a6)--(a7);
\draw[dashed] (a8)--(a4)--(a3);
\draw[dashed] (a1)--(a4);

\node[vertex] at(a1) {};
\node[vertex] at(a2) {};
\node[vertex] at(a3) {};
\node[vertex] at(a4) {};
\node[vertex] at(a5) {};
\node[vertex] at(a6) {};
\node[vertex] at(a7) {};
\node[vertex] at(a8) {};
\node at($(a1)+(t)$) {$1$};
\node at($(a2)+(t)$) {$2$};
\node at($(a3)+(t)$) {$3$};
\node at($(a4)+(t)$) {$4$};
\node at($(a5)+(t)$) {$5$};
\node at($(a6)+(t)$) {$6$};
\node at($(a7)+(t)$) {$7$};
\node at($(a8)+(t)$) {$8$};

\node[vertexe](a12) at($.5*(a1)+.5*(a2)$) {};
\node[vertexe](a23) at($.5*(a2)+.5*(a3)$) {};
\node[vertexe](a34) at($.5*(a3)+.5*(a4)$) {};
\node[vertexe](a41) at($.5*(a4)+.5*(a1)$) {};
\node[vertexe](a15) at($.5*(a1)+.5*(a5)$) {};
\node[vertexe](a26) at($.5*(a2)+.5*(a6)$) {};
\node[vertexe](a37) at($.5*(a3)+.5*(a7)$) {};
\node[vertexe](a48) at($.5*(a4)+.5*(a8)$) {};
\node[vertexe](a56) at($.5*(a5)+.5*(a6)$) {};
\node[vertexe](a67) at($.5*(a6)+.5*(a7)$) {};
\node[vertexe](a78) at($.5*(a7)+.5*(a8)$) {};
\node[vertexe](a85) at($.5*(a8)+.5*(a5)$) {};
\node at($(a12)+(t)$) {\color{black}$9$};
\node at($(a23)+(t)$) {\color{black}$10$};
\node at($(a34)+(t)$) {\color{black}$11$};
\node at($(a41)+(t)$) {\color{black}$12$};
\node at($(a15)+(t)$) {\color{black}$13$};
\node at($(a26)+(t)$) {\color{black}$14$};
\node at($(a37)+(t)$) {\color{black}$15$};
\node at($(a48)+(t)$) {\color{black}$16$};
\node at($(a56)+(t)$) {\color{black}$17$};
\node at($(a67)+(t)$) {\color{black}$18$};
\node at($(a78)+(t)$) {\color{black}$19$};
\node at($(a85)+(t)$) {\color{black}$20$};

\node[vertexf](f1) at($.5*(a12)+.5*(a34)$) {};
\node[vertexf](f2) at($.5*(a12)+.5*(a56)$) {};
\node[vertexf](f3) at($.5*(a23)+.5*(a67)$) {};
\node[vertexf](f4) at($.5*(a34)+.5*(a78)$) {};
\node[vertexf](f5) at($.5*(a41)+.5*(a85)$) {};
\node[vertexf](f6) at($.5*(a56)+.5*(a78)$) {};
\node at($(f1)+(t)$) {\color{gray}$21$};
\node at($(f2)+(t)$) {\color{gray}$22$};
\node at($(f3)+(t)$) {\color{gray}$23$};
\node at($(f4)+(t)$) {\color{gray}$24$};
\node at($(f5)+(t)$) {\color{gray}$25$};
\node at($(f6)+(t)$) {\color{gray}$26$};

\node[vertexv](v) at($.5*(f1)+.5*(f6)$) {};
\node at($(v)+(t)$) {\color{gray}$27$};

\end{tikzpicture}

%% file: figures/tikz/tetrahedron1.tex
\begin{tikzpicture}
\usetikzlibrary{calc}

\makeatletter
\@ifundefined{width}{\newlength\width}{}
\makeatother

\setlength\width{.714\figwidth}

\tikzstyle{line} = [black]
\tikzstyle{tline} = [thick, black]
\tikzstyle{vertex} = [circle, inner sep = 1.4pt, fill = black]

\useasboundingbox (-.1*\width,-.1*(\width) rectangle (1.5*\width,1.5*\width);
\node[coordinate](third) at($(.4*\width,.5*\width)$) {};
\node[coordinate](n1) at($.08*(-\width,-\width)$) {};
\node[coordinate](n2) at($.08*(\width,-\width)$) {};
\node[coordinate](n3) at($.08*(\width,\width)$) {};
\node[coordinate](n4) at($.08*(-\width,\width)$) {};
\node[coordinate](n5) at($.1*(0,-\width)$) {};
\node[coordinate](n6) at($.1*(\width,0)$) {};
\node[coordinate](n7) at($.1*(0,\width)$) {};
\node[coordinate](n8) at($.1*(-\width,0)$) {};

\node[coordinate](a1) at(0, 0) {};
\node[coordinate](a2) at($(a1) + (\width,0)$) {};
\node[coordinate](a3) at($(a2) + (third)$) {};
\node[coordinate](a4) at($(a1) + (third)$) {};
\node[coordinate](a5) at($(a1) + (0,\width)$) {};
\node[coordinate](a6) at($(a5) + (\width,0)$) {};
\node[coordinate](a7) at($(a6) + (third)$) {};
\node[coordinate](a8) at($(a5) + (third)$) {};
\draw[gray] (a5)--(a1)--(a2)--(a6)--(a5)--(a8)--(a7)--(a3)--(a2);
\draw[gray] (a6)--(a7);
\draw[dashed,gray] (a8)--(a4)--(a3);
\draw[dashed,gray] (a1)--(a4);

\draw[thick] (a5)--(a1)--(a2)--cycle;
\draw[thick,dashed] (a5)--(a4)--(a2);
\draw[thick,dashed] (a1)--(a4);

\node[vertex,rectangle,inner sep=1.6pt] at(a1) {};
\node at($(a1)+(n1)$) {$1$};

\end{tikzpicture}

%% file: figures/tikz/tetrahedron9.tex
\begin{tikzpicture}
\usetikzlibrary{calc}

\makeatletter
\@ifundefined{width}{\newlength\width}{}
\makeatother

\setlength\width{.714\figwidth}

\tikzstyle{line} = [black]
\tikzstyle{tline} = [thick, black]
\tikzstyle{vertex} = [circle, inner sep = 1.2pt, fill = black]

\useasboundingbox (-.1*\width,-.1*(\width) rectangle (1.5*\width,1.5*\width);
\node[coordinate](third) at($(.4*\width,.5*\width)$) {};
\node[coordinate](n1) at($.08*(-\width,-\width)$) {};
\node[coordinate](n2) at($.08*(\width,-\width)$) {};
\node[coordinate](n3) at($.08*(\width,\width)$) {};
\node[coordinate](n4) at($.08*(-\width,\width)$) {};
\node[coordinate](n5) at($.1*(0,-\width)$) {};
\node[coordinate](n6) at($.1*(\width,0)$) {};
\node[coordinate](n7) at($.1*(0,\width)$) {};
\node[coordinate](n8) at($.1*(-\width,0)$) {};

\node[coordinate](a1) at(0, 0) {};
\node[coordinate](a2) at($(a1) + (\width,0)$) {};
\node[coordinate](a3) at($(a2) + (third)$) {};
\node[coordinate](a4) at($(a1) + (third)$) {};
\node[coordinate](a5) at($(a1) + (0,\width)$) {};
\node[coordinate](a6) at($(a5) + (\width,0)$) {};
\node[coordinate](a7) at($(a6) + (third)$) {};
\node[coordinate](a8) at($(a5) + (third)$) {};
\draw[gray] (a5)--(a1)--(a2)--(a6)--(a5)--(a8)--(a7)--(a3)--(a2);
\draw[gray] (a6)--(a7);
\draw[dashed,gray] (a8)--(a4)--(a3);
\draw[dashed,gray] (a1)--(a4);

\node[coordinate](a12) at($.5*(a1)+.5*(a2)$) {};
\node[coordinate](a23) at($.5*(a2)+.5*(a3)$) {};
\node[coordinate](a34) at($.5*(a3)+.5*(a4)$) {};
\node[coordinate](a41) at($.5*(a4)+.5*(a1)$) {};
\node[coordinate](a15) at($.5*(a1)+.5*(a5)$) {};
\node[coordinate](a26) at($.5*(a2)+.5*(a6)$) {};
\node[coordinate](a37) at($.5*(a3)+.5*(a7)$) {};
\node[coordinate](a48) at($.5*(a4)+.5*(a8)$) {};
\node[coordinate](a56) at($.5*(a5)+.5*(a6)$) {};
\node[coordinate](a67) at($.5*(a6)+.5*(a7)$) {};
\node[coordinate](a78) at($.5*(a7)+.5*(a8)$) {};
\node[coordinate](a85) at($.5*(a8)+.5*(a5)$) {};

\draw[thick] (a1)--(a2)--(a56)--cycle;
\draw[thick,dashed] (a1)--(a34);
\draw[thick] (a56)--(a34)--(a2);

\node[vertex,rectangle,inner sep=2pt] at(a12) {};
\node at($(a12)+(n5)$) {$9$};

\end{tikzpicture}

%% file: figures/tikz/tetrahedron2.tex
\begin{tikzpicture}
\usetikzlibrary{calc}

\makeatletter
\@ifundefined{width}{\newlength\width}{}
\makeatother

\setlength\width{.714\figwidth}

\tikzstyle{line} = [black]
\tikzstyle{tline} = [thick, black]
\tikzstyle{vertex} = [circle, inner sep = 1.4pt, fill = black]

\useasboundingbox (-.1*\width,-.1*(\width) rectangle (1.5*\width,1.5*\width);
\node[coordinate](third) at($(.4*\width,.5*\width)$) {};
\node[coordinate](n1) at($.08*(-\width,-\width)$) {};
\node[coordinate](n2) at($.08*(\width,-\width)$) {};
\node[coordinate](n3) at($.08*(\width,\width)$) {};
\node[coordinate](n4) at($.08*(-\width,\width)$) {};
\node[coordinate](n5) at($.1*(0,-\width)$) {};
\node[coordinate](n6) at($.1*(\width,0)$) {};
\node[coordinate](n7) at($.1*(0,\width)$) {};
\node[coordinate](n8) at($.1*(-\width,0)$) {};

\node[coordinate](a1) at(0, 0) {};
\node[coordinate](a2) at($(a1) + (\width,0)$) {};
\node[coordinate](a3) at($(a2) + (third)$) {};
\node[coordinate](a4) at($(a1) + (third)$) {};
\node[coordinate](a5) at($(a1) + (0,\width)$) {};
\node[coordinate](a6) at($(a5) + (\width,0)$) {};
\node[coordinate](a7) at($(a6) + (third)$) {};
\node[coordinate](a8) at($(a5) + (third)$) {};
\draw[gray] (a5)--(a1)--(a2)--(a6)--(a5)--(a8)--(a7)--(a3)--(a2);
\draw[gray] (a6)--(a7);
\draw[dashed,gray] (a8)--(a4)--(a3);
\draw[dashed,gray] (a1)--(a4);

\draw[thick] (a6)--(a1)--(a2)--(a3)--cycle;
\draw[thick] (a6)--(a2);
\draw[thick,dashed] (a1)--(a3);

\node[vertex,rectangle,inner sep=2pt] at(a2) {};
\node at($(a2)+(n2)$) {$2$};

\end{tikzpicture}

%% file: figures/tikz/tetrahedron10.tex
\begin{tikzpicture}
\usetikzlibrary{calc}

\makeatletter
\@ifundefined{width}{\newlength\width}{}
\makeatother

\setlength\width{.714\figwidth}

\tikzstyle{line} = [black]
\tikzstyle{tline} = [thick, black]
\tikzstyle{vertex} = [circle, inner sep = 1.2pt, fill = black]

\useasboundingbox (-.1*\width,-.1*(\width) rectangle (1.5*\width,1.5*\width);
\node[coordinate](third) at($(.4*\width,.5*\width)$) {};
\node[coordinate](n1) at($.08*(-\width,-\width)$) {};
\node[coordinate](n2) at($.08*(\width,-\width)$) {};
\node[coordinate](n3) at($.08*(\width,\width)$) {};
\node[coordinate](n4) at($.08*(-\width,\width)$) {};
\node[coordinate](n5) at($.1*(0,-\width)$) {};
\node[coordinate](n6) at($.1*(\width,0)$) {};
\node[coordinate](n7) at($.1*(0,\width)$) {};
\node[coordinate](n8) at($.1*(-\width,0)$) {};

\node[coordinate](a1) at(0, 0) {};
\node[coordinate](a2) at($(a1) + (\width,0)$) {};
\node[coordinate](a3) at($(a2) + (third)$) {};
\node[coordinate](a4) at($(a1) + (third)$) {};
\node[coordinate](a5) at($(a1) + (0,\width)$) {};
\node[coordinate](a6) at($(a5) + (\width,0)$) {};
\node[coordinate](a7) at($(a6) + (third)$) {};
\node[coordinate](a8) at($(a5) + (third)$) {};
\draw[gray] (a5)--(a1)--(a2)--(a6)--(a5)--(a8)--(a7)--(a3)--(a2);
\draw[gray] (a6)--(a7);
\draw[dashed,gray] (a8)--(a4)--(a3);
\draw[dashed,gray] (a1)--(a4);

\node[coordinate](a12) at($.5*(a1)+.5*(a2)$) {};
\node[coordinate](a23) at($.5*(a2)+.5*(a3)$) {};
\node[coordinate](a34) at($.5*(a3)+.5*(a4)$) {};
\node[coordinate](a41) at($.5*(a4)+.5*(a1)$) {};
\node[coordinate](a15) at($.5*(a1)+.5*(a5)$) {};
\node[coordinate](a26) at($.5*(a2)+.5*(a6)$) {};
\node[coordinate](a37) at($.5*(a3)+.5*(a7)$) {};
\node[coordinate](a48) at($.5*(a4)+.5*(a8)$) {};
\node[coordinate](a56) at($.5*(a5)+.5*(a6)$) {};
\node[coordinate](a67) at($.5*(a6)+.5*(a7)$) {};
\node[coordinate](a78) at($.5*(a7)+.5*(a8)$) {};
\node[coordinate](a85) at($.5*(a8)+.5*(a5)$) {};

\draw[thick] (a2)--(a3)--(a67)--(a41)--cycle;
\draw[thick] (a2)--(a67);
\draw[thick,dashed] (a41)--(a3);

\node[vertex,rectangle,inner sep=2pt] at(a23) {};
\node at($(a23)+(n6)$) {$10$};

\end{tikzpicture}

%% file: main_hexValidity.bbl
\begin{thebibliography}{24}
\expandafter\ifx\csname natexlab\endcsname\relax\def\natexlab#1{#1}\fi
\providecommand{\url}[1]{\texttt{#1}}
\providecommand{\href}[2]{#2}
\providecommand{\path}[1]{#1}
\providecommand{\DOIprefix}{doi:}
\providecommand{\ArXivprefix}{arXiv:}
\providecommand{\URLprefix}{URL: }
\providecommand{\Pubmedprefix}{pmid:}
\providecommand{\doi}[1]{\href{http://dx.doi.org/#1}{\path{#1}}}
\providecommand{\Pubmed}[1]{\href{pmid:#1}{\path{#1}}}
\providecommand{\bibinfo}[2]{#2}
\ifx\xfnm\relax \def\xfnm[#1]{\unskip,\space#1}\fi
\bibitem[{Johnen et~al.(2013)Johnen, Remacle, and
  Geuzaine}]{johnen2013geometrical}
\bibinfo{author}{A.~Johnen}, \bibinfo{author}{J.-F. Remacle},
  \bibinfo{author}{C.~Geuzaine},
\newblock \bibinfo{title}{Geometrical validity of curvilinear finite elements},
\newblock \bibinfo{journal}{Journal of Computational Physics}
  \bibinfo{volume}{233} (\bibinfo{year}{2013}) \bibinfo{pages}{359--372}.
\bibitem[{Remacle et~al.(2016)Remacle, Gandham, and Warburton}]{remacle2016gpu}
\bibinfo{author}{J.-F. Remacle}, \bibinfo{author}{R.~Gandham},
  \bibinfo{author}{T.~Warburton},
\newblock \bibinfo{title}{Gpu accelerated spectral finite elements on all-hex
  meshes},
\newblock \bibinfo{journal}{Journal of Computational Physics}
  \bibinfo{volume}{324} (\bibinfo{year}{2016}) \bibinfo{pages}{246--257}.
\bibitem[{Wang et~al.(2004)Wang, Nelson, and Rauch}]{wang2004back}
\bibinfo{author}{E.~Wang}, \bibinfo{author}{T.~Nelson},
  \bibinfo{author}{R.~Rauch},
\newblock \bibinfo{title}{Back to elements—tetrahedra vs. hexahedra},
\newblock in: \bibinfo{booktitle}{Proceedings of the 2004 International ANSYS
  Conference}, \bibinfo{organization}{ANSYS Pennsylvania},
  \bibinfo{year}{2004}.
\bibitem[{Benzley et~al.(1995)Benzley, Perry, Merkley, Clark, and
  Sjaardama}]{benzley1995comparison}
\bibinfo{author}{S.~E. Benzley}, \bibinfo{author}{E.~Perry},
  \bibinfo{author}{K.~Merkley}, \bibinfo{author}{B.~Clark},
  \bibinfo{author}{G.~Sjaardama},
\newblock \bibinfo{title}{A comparison of all-hexagonal and all-tetrahedral
  finite element meshes for elastic and elasto-plastic analysis},
\newblock in: \bibinfo{booktitle}{Proceedings of the 4th International Meshing
  Roundtable}, volume~\bibinfo{volume}{17}, \bibinfo{organization}{Sandia
  National Laboratories Albuquerque, NM}, \bibinfo{year}{1995}, pp.
  \bibinfo{pages}{179--191}.
\bibitem[{Baudouin et~al.(2014)Baudouin, Remacle, Marchandise, Henrotte, and
  Geuzaine}]{baudouin2014frontal}
\bibinfo{author}{T.~C. Baudouin}, \bibinfo{author}{J.-F. Remacle},
  \bibinfo{author}{E.~Marchandise}, \bibinfo{author}{F.~Henrotte},
  \bibinfo{author}{C.~Geuzaine},
\newblock \bibinfo{title}{A frontal approach to hex-dominant mesh generation},
\newblock \bibinfo{journal}{Advanced Modeling and Simulation in Engineering
  Sciences} \bibinfo{volume}{1} (\bibinfo{year}{2014}) \bibinfo{pages}{1--30}.
\bibitem[{Botella et~al.(2016)Botella, L{\'e}vy, and
  Caumon}]{botella2016indirect}
\bibinfo{author}{A.~Botella}, \bibinfo{author}{B.~L{\'e}vy},
  \bibinfo{author}{G.~Caumon},
\newblock \bibinfo{title}{Indirect unstructured hex-dominant mesh generation
  using tetrahedra recombination},
\newblock \bibinfo{journal}{Computational Geosciences} \bibinfo{volume}{20}
  (\bibinfo{year}{2016}) \bibinfo{pages}{437--451}.
\bibitem[{Sokolov et~al.(2016)Sokolov, Ray, Untereiner, and
  L{\'e}vy}]{sokolov2016hexahedral}
\bibinfo{author}{D.~Sokolov}, \bibinfo{author}{N.~Ray},
  \bibinfo{author}{L.~Untereiner}, \bibinfo{author}{B.~L{\'e}vy},
\newblock \bibinfo{title}{Hexahedral-dominant meshing},
\newblock \bibinfo{journal}{ACM Transactions on Graphics (TOG)}
  \bibinfo{volume}{35} (\bibinfo{year}{2016}) \bibinfo{pages}{157}.
\bibitem[{Pellerin et~al.(2017)Pellerin, Johnen, and Remacle}]{pellerin2017}
\bibinfo{author}{J.~Pellerin}, \bibinfo{author}{A.~Johnen},
  \bibinfo{author}{J.-F. Remacle},
\newblock \bibinfo{title}{Identifying combinations of tetrahedra into
  hexahedra: a vertex based strategy},
\newblock in: \bibinfo{booktitle}{Proceedings of the 26th International Meshing
  Roundtable}, \bibinfo{year}{2017}.
\bibitem[{Knupp(1990)}]{knupp1990invertibility}
\bibinfo{author}{P.~M. Knupp},
\newblock \bibinfo{title}{On the invertibility of the isoparametric map},
\newblock \bibinfo{journal}{Computer Methods in Applied Mechanics and
  Engineering} \bibinfo{volume}{78} (\bibinfo{year}{1990})
  \bibinfo{pages}{313--329}.
\bibitem[{Ivanenko(1999)}]{ivanenko1999harmonic}
\bibinfo{author}{S.~A. Ivanenko},
\newblock \bibinfo{title}{Harmonic mappings},
\newblock in: \bibinfo{booktitle}{Handbook of grid generation},
  \bibinfo{publisher}{CRC Press Boca Raton, Fl}, \bibinfo{year}{1999}.
\bibitem[{Grandy(1999)}]{grandy1999conservative}
\bibinfo{author}{J.~Grandy},
\newblock \bibinfo{title}{Conservative remapping and region overlays by
  intersecting arbitrary polyhedra},
\newblock \bibinfo{journal}{Journal of Computational Physics}
  \bibinfo{volume}{148} (\bibinfo{year}{1999}) \bibinfo{pages}{433--466}.
\bibitem[{Ushakova(2001)}]{ushakova2001conditions}
\bibinfo{author}{O.~V. Ushakova},
\newblock \bibinfo{title}{Conditions of nondegeneracy of three-dimensional
  cells. a formula of a volume of cells},
\newblock \bibinfo{journal}{SIAM Journal on Scientific Computing}
  \bibinfo{volume}{23} (\bibinfo{year}{2001}) \bibinfo{pages}{1274--1290}.
\bibitem[{Vavasis(2003)}]{vavasis2003bernstein}
\bibinfo{author}{S.~Vavasis}, \bibinfo{title}{A bernstein-bezier sufficient
  condition for invertibility of polynomial mapping functions},
  \bibinfo{year}{2003}. \bibinfo{note}{Draft}.
\bibitem[{Shangyou(2005)}]{shangyou2005subtetrabedral}
\bibinfo{author}{Z.~Shangyou}, \bibinfo{title}{Subtetrabedral test for the
  positive jacobian of hexaherdral elements}, \bibinfo{year}{2005}.
  \bibinfo{note}{Unpublished}.
\bibitem[{Ushakova(2011)}]{ushakova2011nondegeneracy}
\bibinfo{author}{O.~V. Ushakova},
\newblock \bibinfo{title}{Nondegeneracy tests for hexahedral cells},
\newblock \bibinfo{journal}{Computer Methods in Applied Mechanics and
  Engineering} \bibinfo{volume}{200} (\bibinfo{year}{2011})
  \bibinfo{pages}{1649--1658}.
\bibitem[{Knabner et~al.(2003)Knabner, Korotov, and
  Summ}]{knabner2003conditions}
\bibinfo{author}{P.~Knabner}, \bibinfo{author}{S.~Korotov},
  \bibinfo{author}{G.~Summ},
\newblock \bibinfo{title}{Conditions for the invertibility of the isoparametric
  mapping for hexahedral finite elements},
\newblock \bibinfo{journal}{Finite elements in analysis and design}
  \bibinfo{volume}{40} (\bibinfo{year}{2003}) \bibinfo{pages}{159--172}.
\bibitem[{Geuzaine and Remacle(2009)}]{geuzaine2009gmsh}
\bibinfo{author}{C.~Geuzaine}, \bibinfo{author}{J.-F. Remacle},
\newblock \bibinfo{title}{Gmsh: A {3-D} finite element mesh generator with
  built-in pre-and post-processing facilities},
\newblock \bibinfo{journal}{International Journal for Numerical Methods in
  Engineering} \bibinfo{volume}{79} (\bibinfo{year}{2009})
  \bibinfo{pages}{1309--1331}.
\bibitem[{Frey et~al.(1978)Frey, Hall, and Porsching}]{frey1978some}
\bibinfo{author}{A.~E. Frey}, \bibinfo{author}{C.~A. Hall},
  \bibinfo{author}{T.~A. Porsching},
\newblock \bibinfo{title}{Some results on the global inversion of bilinear and
  quadratic isoparametric finite element transformations},
\newblock \bibinfo{journal}{Mathematics of Computation} \bibinfo{volume}{32}
  (\bibinfo{year}{1978}) \bibinfo{pages}{725--749}.
\bibitem[{Zhang(2005)}]{zhang2005bijectivity}
\bibinfo{author}{S.~Zhang}, \bibinfo{title}{Invertible jacobian for hexahedral
  finite elements. part 1. bijectivity},
  \bibinfo{howpublished}{\url{http://www.math.udel.edu/\~szhang/research/p/bijective1.ps}},
  \bibinfo{year}{2005}.
\bibitem[{Leroy(2008)}]{leroy2008certificats}
\bibinfo{author}{R.~Leroy}, \bibinfo{title}{Certificats de positivit\'{e} et
  minimisation polynomiale dans la base de {B}ernstein multivari\'{e}e}, Ph.D.
  thesis, Universit\'{e} de Rennes 1, \bibinfo{year}{2008}.
\bibitem[{Leroy(2011)}]{leroy2011certificats}
\bibinfo{author}{R.~Leroy}, \bibinfo{title}{Certificates of positivity in the
  simplicial bernstein basis},
  \bibinfo{howpublished}{\url{https://hal.archives-ouvertes.fr/hal-00589945/document}},
  \bibinfo{year}{2011}.
\bibitem[{Cohen and Schumaker(1985)}]{cohen1985rates}
\bibinfo{author}{E.~Cohen}, \bibinfo{author}{L.~L. Schumaker},
\newblock \bibinfo{title}{Rates of convergence of control polygons},
\newblock \bibinfo{journal}{Computer Aided Geometric Design}
  \bibinfo{volume}{2} (\bibinfo{year}{1985}) \bibinfo{pages}{229--235}.
\bibitem[{Knupp(2000)}]{knupp2000achievingB}
\bibinfo{author}{P.~M. Knupp},
\newblock \bibinfo{title}{Achieving finite element mesh quality via
  optimization of the {Jacobian} matrix norm and associated quantities. part
  {II}---{A} framework for volume mesh optimization and the condition number of
  the jacobian matrix},
\newblock \bibinfo{journal}{International Journal for Numerical Methods in
  Engineering} \bibinfo{volume}{48} (\bibinfo{year}{2000})
  \bibinfo{pages}{1165--1185}.
\bibitem[{Yamakawa and Shimada(2003)}]{yamakawa2003fully}
\bibinfo{author}{S.~Yamakawa}, \bibinfo{author}{K.~Shimada},
\newblock \bibinfo{title}{Fully-automated hex-dominant mesh generation with
  directionality control via packing rectangular solid cells},
\newblock \bibinfo{journal}{International Journal for Numerical Methods in
  Engineering} \bibinfo{volume}{57} (\bibinfo{year}{2003})
  \bibinfo{pages}{2099--2129}.

\end{thebibliography}
